\documentclass[amsmath,amssymb,aps,pra,twocolumn, superscriptaddress,longbibliography]{revtex4-2}

\usepackage{graphicx}
\usepackage{dcolumn}
\usepackage{xcolor}

\usepackage{helvet}
\usepackage{mathrsfs}
\usepackage[colorlinks=true,bookmarks=false,citecolor=blue,urlcolor=blue]{hyperref}
\usepackage{physics}
\usepackage{dsfont}

\newcommand{\<}{\langle}
\renewcommand{\>}{\rangle}

\newcommand{\blue}[1]{\textcolor{black}{#1}}

\renewcommand{\cite}[1]{[\onlinecite{#1}]}
\newcommand{\cra}[1]{\hat{a}^{\dag}_{#1}}  
\newcommand{\ana}[1]{\hat{a}_{#1}}         

\newcommand{\df}[2]{\frac{\partial #1}{\partial #2}} 

\begin{document}

\title{Efficient computation of quantum time-optimal control}

\author{Andrei~A.~Stepanenko}
\email{as@lims.ac.uk}
\affiliation{London Institute for Mathematical Sciences, Royal Institution, London, UK}
\affiliation{School of Physics and Engineering, ITMO University, Saint Petersburg 197101, Russia}

\author{Kseniia S. Chernova}
\affiliation{School of Physics and Engineering, ITMO University, Saint Petersburg 197101, Russia}

\author{Maxim Mazanov}
\affiliation{School of Physics and Engineering, ITMO University, Saint Petersburg 197101, Russia}

\author{Maxim~A.~Gorlach}
\email{m.gorlach@metalab.ifmo.ru}
\affiliation{School of Physics and Engineering, ITMO University, Saint Petersburg 197101, Russia}

\begin{abstract}
We present an approach to compute time-optimal control of a quantum system which combines quantum brachistochrone and Lax pair techniques and enables efficient investigation of large-scale quantum systems. We illustrate our method by finding the \blue{quantum speed limit for} a single-particle excitation in a nearest-neighbor-coupled qubit lattice with \blue{switchable couplings} and fixed sum of their squares. \blue{We obtain the solution for a finite system with up to 10 000 qubits and for the effectively infinite lattice closed into ring. As another application of our method, we derive the quantum speed limit for a lattice with time-independent couplings.}
\end{abstract}


\maketitle

{\it Introduction.~}Harnessing the strengths and versatility of quantum systems requires their flexible and comprehensive control. The relevant strategies are explored by the quantum optimal control theory~\cite{Werschnik2007,Rabitz2010,Boscain2021,Ansel2024} and allow one to boost the performance of quantum systems in an experimental situation~\cite{Geng2016,Lam2021,Chen2025}. One of the important goals in this field is time-optimal quantum evolution, which brings the system to the desired quantum state within the minimal possible time.

A promising approach to find such quantum time-optimal control is the quantum brachistochrone technique~\cite{Carlini2006,Carlini2007,Carlini2008,Wang2015,Koike2022,Malikis2024}, which is the variational method minimizing the evolution time given the prescribed constraints on the system Hamiltonian.

Until recently, the solutions were available only for the relatively small  systems~\cite{Carlini2011,Carlini2012,Carlini2013,Carlini2017,Wu2024,Chernova2025,WangePRL2025,Himmel2026}. At the same time, it is well understood that the full power of quantum physics can be unraveled only in the large-scale arrays of qubits~\cite{GoogleQuantumAI,GoogleQuantumAI-Oct2025,Chiu2025} and hence finding the pathways to optimally control such systems is of paramount importance. 

The main challenge of examining large-scale quantum systems lies in the rapid growth of the number of unknown variables with the size of the system. Specifically, if the Hilbert space of the quantum system is $N$-dimensional, one needs to find not only the $N$-component wave function, but also all controls~-- the elements of the Hamiltonian. This yields $\propto N^2$ unknowns in the quantum brachistochrone method, reducing to $\propto N^2/2$ in some cases~\cite{Stepanenko2025}.

In this Letter, we make a conceptual advance in finding time-optimal control of large-scale quantum systems. We show that the number of unknown variables can be reduced to $2\,N$ and even $N$ in the special instances, which greatly decreases the complexity of the control problem. Our key idea is to introduce a pair of complex-valued auxiliary $N$-component vectors. While being abstract, these vectors fully encode not only the wave function $\ket{\psi}$, but also the system Hamiltonian $\hat{H}$. This finding, in turn, drastically simplifies the control equations and their solution. 

{\it Methodology.~}The starting point of our approach is the quantum brachistochrone technique for the conservative case~\cite{Carlini2006,Carlini2007,Wang2015,Koike2022,Malikis2024,Stepanenko2025}. The wave function $\ket{\psi}$ \blue{evolves according to the Schr\"odinger equation}
%
\begin{equation}\label{eq:Schr}
    \partial_t\ket{\psi}=-i\,\hat{H}\,\ket{\psi}\:.
\end{equation}
%
We minimize the cost functional
\begin{gather}
    S = \frac{1}{2}\,\int_0^\tau \text{Tr}\,\hat{H}^2\, dt+\int_0^\tau \text{Tr}\left(\hat{D} \hat{H}\right) dt \notag\\  +\int_0^\tau\text{Tr}\left(\hat{R}_t(\partial_t\hat{\rho} + i[\hat{H},\hat{\rho}])\right)\,dt\:.\label{eq:Cost}
\end{gather}
The first term in Eq.~\eqref{eq:Cost} is introduced to minimize the norm of the Hamiltonian $\|\hat{H}\|=\sqrt{\text{Tr}\,\hat{H}^2}$ for the fixed evolution time $\tau=1$. This is equivalent to minimizing the evolution time $\tau$ for the fixed $\|\hat{H}\|^2$. In addition, the result does not depend on whether we minimize  $\int_0^\tau \|\hat{H}\|\,dt$ or $1/2\,\int_0^\tau \|\hat{H}\|^2 dt$, see Supplementary Materials~\cite{Supplement}, Sec.~I.

The second term in Eq.~\eqref{eq:Cost} restricts the structure of the Hamiltonian by the conditions $\text{Tr}(\hat{H}\hat{D}_k)=0$, where $\hat{D}=\sum\limits_k\,\lambda_k\,\hat{D}_k$ and $\lambda_k$ are unknown time-dependent Lagrange multipliers. This captures physical constraints on the system Hamiltonian dictated by the experimental setting. \blue{For instance, this term prohibits complex and long-range couplings in a one-dimensional tight-binding lattice.}


\blue{The third term with Lagrange multiplier matrix $\hat{R}$ ensures that the evolution of the density matrix $\hat{\rho}(t)=\ket{\psi(t)}\bra{\psi(t)}$ is unitary and satisfies the equation
\begin{equation}\label{eq:Rho}
\partial_t\hat{\rho}+i[\hat{H},\hat{\rho}]=0\:.
\end{equation}
} Note that the initial and final states of the system are fixed up to the global phase, i.e. $\delta\hat{\rho}_0=\delta\hat{\rho}_\tau=0$. 
%

\blue{Next we define the operator $\hat{L}$ as}
\begin{equation}\label{eq:QBE_bb}
\hat{L}(t) = i\left[\hat{R}(t),\hat{\rho}(t)\right].
\end{equation}
\blue{By varying the cost functional with respect to the dynamical variables $\hat{H}$ and $\hat{\rho}$~\cite{Supplement} (Sec.~I), we derive}
\begin{gather}
\hat{L}=\hat{H}+\hat{D}\:,\label{eq:LaxResult}\\
\partial_t\hat{R}+i[\hat{H},\hat{R}]=0\:.\label{eq:Revol}
\end{gather}
Combining Eqs.~\eqref{eq:Rho},\eqref{eq:QBE_bb},\eqref{eq:Revol}, we recover quantum brachistochrone equation
\begin{equation} \label{eq:QBE}
\partial_t \hat{L} = -i\left[\hat{H},\hat{L}\right] .  
\end{equation}
%
%
%
The Hermitian matrix $\hat{R}$ of Lagrange multipliers can be excluded from Eq.~\eqref{eq:QBE_bb} to yield
\begin{equation}\label{eq:BC1}
\hat{\rho}\, \hat{L}(t)\, \hat{\rho}  = 0, \mspace{6mu}   \left(\hat{I}-\hat{\rho}\right) \hat{L}(t) \left(\hat{I}-\hat{\rho}\right)  = 0\:.
\end{equation}
These results suggest a deep connection between the evolution of the operator $\hat{L}$ and the wave function $\ket{\psi}$. This connection can be further formalized by observing that $-i\hat{H}$ and $\hat{L}$ operators form {\it the Lax pair}~\cite{Babelon2003} and that the eigenvalues $\gamma$ of the Lax operator $\hat{L}$ are time-independent (Ref.~\cite{Babelon2003} and  \cite{Supplement}, Sec.~II). Equation~\eqref{eq:QBE_bb} further simplifies the Lax operator and ensures only two nonzero eigenvalues with the opposite signs 
at the arbitrary moment of time (Ref.~\cite{Supplement}, Sec.~II)
\begin{equation}\label{eq:Lax}
 \hat{L}(t) = L_0\ket{\alpha(t)}\bra{\alpha(t)} -L_0 \ket{\beta(t)}\bra{\beta(t)}\:,   
\end{equation}
where Lax eigenvectors $\ket{\alpha(t)}$ and $\ket{\beta(t)}$ evolve according to Eq.~\eqref{eq:Schr}, are normalized to unity: $\<\alpha|\alpha\>=\<\beta|\beta\>=1$, being orthogonal to each other: $\<\alpha|\beta\>=0$.

Crucially, if one knows the vectors $\ket{\alpha(t)}$ and $\ket{\beta(t)}$, the Hamiltonian of the system is recovered from the Lax operator Eq.~\eqref{eq:Lax} by projecting out known $\hat{D}_k$ matrices. The same vectors define the state of the quantum system via~\cite{Supplement} (Sec.~II)
\begin{equation}
    \ket{\psi(t)}=\frac{1}{\sqrt{2}}\,\left(\ket{\alpha(t)}+\ket{\beta(t)}\right)\:.
\end{equation}
Finally, the initial and boundary conditions for the vector $\ket{\psi}$ translate into the conditions for $\ket{\alpha}$ and $\ket{\beta}$ vectors.

Our approach thus reduces a complicated quantum optimal control problem with the unknown wave function, Hamiltonian and a set of Lagrange multipliers to finding the evolution of just two $N$-component vectors, $\ket{\alpha(t)}$ and $\ket{\beta(t)}$. The above formulation is fully general and independent of the specific Hamiltonian, initial and final quantum states and only assumes a unitary evolution. Moreover, in special cases there is a simple connection between the vectors $\ket{\alpha(t)}$ and $\ket{\beta(t)}$, which reduces the problem to just $N$ unknown complex amplitudes~-- the same amount as in the problem with the time-independent Hamiltonian.



\begin{figure}[b]
    \centering
    \includegraphics[width=0.6\linewidth]{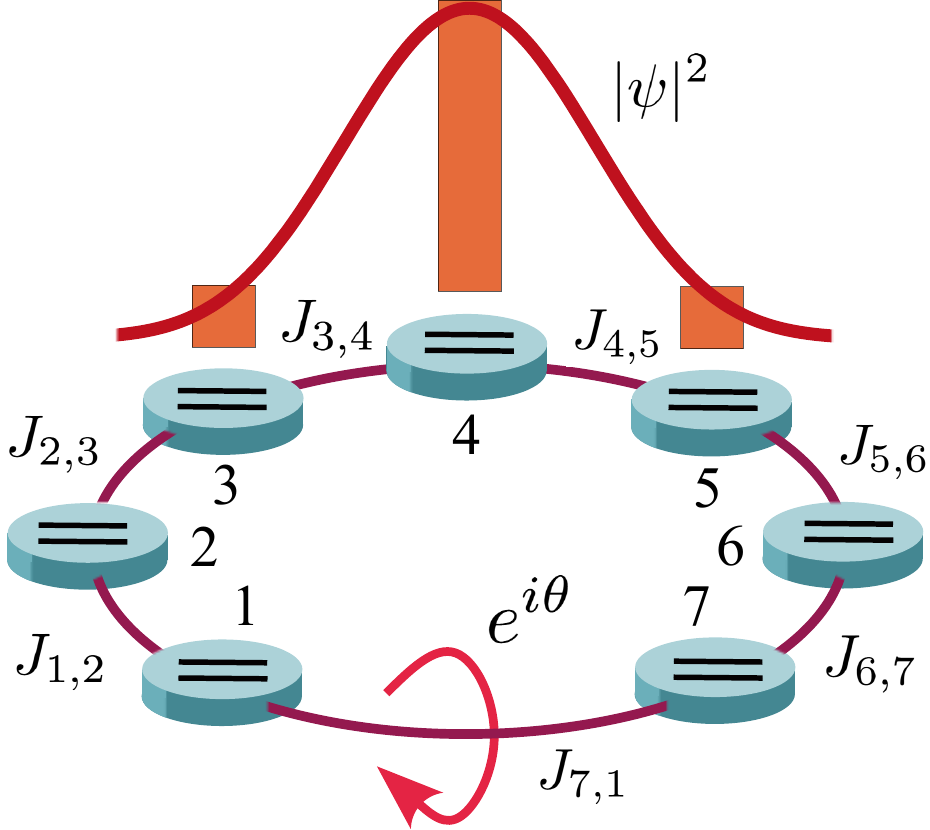}
    \caption{Schematic of a qubit array with \blue{controllable couplings}, imposed anti-periodic boundary conditions and different number $N$ of qubits comprising the ring. Quantum speed limit is achieved for the excitation propagating in the form of a localized wave packet maintaining its shape.}
    \label{fig:1}
\end{figure}

{\it Applications.~}As an illustration of our method, we investigate the transfer of a single-particle excitation in a qubit array with the nearest-neighbor couplings controlled in time~\cite{Yan2018,Chen2014}, Fig.~\ref{fig:1}. We also assume that the overall magnitude of the couplings is restricted by the bound $\sum_m\,J_{m,m+1}^2=J_0^2$. Intuitively, $J_0$ quantifies the available resource which can be distributed between the different coupling links in order to speed up the transfer. While the specific form of the constraint may look artificial, this streamlines the theoretical analysis and highlights fruitful parallels with the geodesic problem in geometry~\cite{Anandan1990,Wang2015,Meinersen2024}. 

We study the problem of time-optimal transfer in the absence of dissipation, which is a realistic assumption for the arrays of superconducting qubits where the coherence time ranges from tens and hundreds of microseconds to a millisecond~\cite{Somoroff2023} exceeding the duration of the standard gates by orders of magnitude.



The Hamiltonian of the described array of $N$ qubits with the real-valued nearest-neighbor couplings $J_{m,m+1}(t)$ [Fig.~\ref{fig:1}] reads
%
%
\begin{equation}
    \label{eq:Ham}
    \hat{H}(t) = \sum_{m=1}^{N} \left(J_{m,m+1}(t)\, \cra{m}\ana{m+1}+\text{H.c.}\right)\,,
\end{equation}
where $\hbar = 1$ and $\cra{m}$, $\ana{m}$ are creation and annihilation bosonic operators. The eigenfrequencies of all qubits are assumed equal to each other and set to zero for clarity. The system is made effectively infinite by closing the array into ring and identifying $J_{N,N+1}\equiv J_{N,1}$.


The Hamiltonian Eq.~\eqref{eq:Ham} commutes with the total number of excitations. Since we study the single-particle dynamics, the relevant sector of the Hilbert space is spanned by the vectors $\ket{m}=\cra{m}\,\ket{0}$ and has the dimensionality $N$ equal to the number of qubits. In such representation, the Hamiltonian is $N\times N$ traceless Hermitian matrix with the only nonzero elements $H_{m,m+1} = H_{m+1,m} = J_{m,m+1}$ and $H_{1,N} = H_{N,1} = J_{1,N}$.


The space $\mathcal{M}$ of traceless Hermitian matrices with the dimensionality $N^2-1$ can be decomposed into  two subspaces $\mathcal{A}$ and $\mathcal{B}$ such that $\mathcal{A}\cup\mathcal{B}=\mathcal{M}$. The subspace $\mathcal{A}$ is spanned by the matrices
%
\begin{equation}
 \hat{A}_{m,m+p} = \frac{1}{\sqrt{2}}\,(i^{p-1}\ket{m}\bra{m+p}+i^{1-p}\ket{m+p}\bra{m}) 
\end{equation}
with $1\leq m\leq N-1$ and $1\leq p\leq N-m$. The dimensionality of this subspace is $N(N-1)/2$ and the Hamiltonian is presented as $\hat{H} = \sqrt{2}\, \sum_m J_{m,m+1}\,\hat{A}_{m,m+1}$. The remaining $(N-1)(N+2)/2$ matrices $\hat{B}_l$ form the basis in the subspace $\mathcal{B}$. All introduced basis matrices are orthogonal and normalized by the conditions $\text{Tr}(\hat{A}_{m,n}\hat{A}_{p,q}) = \delta_{mp}\delta_{nq}$, $\text{Tr}(\hat{B}_{l,k}\hat{B}_{p,q}) = \delta_{lp}\delta_{kq}$, $\text{Tr}(\hat{A}_{m,n}\hat{B}_{l,k}) = 0$.



The subspace $\mathcal{A}$ is closed under the commutation $i[\mathcal{A},\mathcal{A}] \subseteq \mathcal{A}$, while $i[\mathcal{B},\mathcal{A}]\subseteq \mathcal{B}$. As shown in Ref.~\cite{Supplement}, Sec.~III, this  decouples the Lagrange multipliers from the $\mathcal{B}$ subspace, and the dynamics of interest happens only in the $\mathcal{A}$ subspace.


Further analysis is largely simplified by performing a unitary transformation $\hat{Q}$ with $Q_{mm}=i^{m-1}$ which renders the Hamiltonian and the Lax operator projected onto $\mathcal{A}$ subspace, $\hat{L}^A$, purely imaginary. In this representation, Lax eigenvectors $\ket{a}=\hat{Q}\ket{\alpha}$ and $\ket{b}=\hat{Q}\ket{\beta}$ are connected to each other in a  simple way $\ket{b}=\ket{a^*}$ (\cite{Supplement}, Sec.~III), which yields the wave function 
\begin{equation}\label{eq:WaveFunc}
 \ket{\psi^Q}=(\ket{a}+\ket{a^*})/\sqrt{2}   
\end{equation}
and the Hamiltonian elements
\begin{equation}\label{eq:Hq-elements}
\begin{split}
 H_{m,m\pm1}^Q=L_{m,m\pm1}^Q=L_0\,(\ket{a}\bra{a}-\ket{a^*}\bra{a^*})_{m,m\pm1}\\
 =L_0\,(a_m a_{m\pm1}^*-a_m^*\,a_{m\pm1})\:.
\end{split}
\end{equation}
Using the expression for $\hat{H}^Q$ in the evolution equation for the Lax eigenvector $i\,\partial_t\ket{a}=\hat{H}^Q\ket{a}$, we finally recover a set of nonlinear differential equations for the components of the Lax eigenvector:
\begin{eqnarray}\label{eq:Lax_QBE_a}
\frac{i}{L_0}\,\df{a_m}{t} &=& \left(|a_{m-1}|^2+|a_{m+1}|^2\right)\,a_m\nonumber\\&&
-\left(a_{m-1}^2+a_{m+1}^2\right)\,a_m^*\:,
\end{eqnarray}
where $1\leq m\leq N$, Lax eigenvector is normalized as $\<a|a\>=1$, $L_0$ is a real positive constant and appropriate boundary conditions have to be imposed.

It should be stressed that the system~\eqref{eq:Lax_QBE_a} contains only $N$ scalar equations for the complex amplitudes $a_m$. Nevertheless, they encode full dynamics defining both the evolution of the wave function and the Hamiltonian via Eqs.~\eqref{eq:WaveFunc}, \eqref{eq:Hq-elements}. As a result, the complexity of time-optimal control problem becomes comparable to the case of time-independent Hamiltonians which opens an avenue to the efficient simulation of large-scale optimally controlled quantum systems.

Effective nonlinearity arising in Eq.~\eqref{eq:Lax_QBE_a} is due to the self-consistent nature of our problem when the Hamiltonian changes synchronously with the evolution of the quantum state. Previously, such discrete nonlinear equations were studied in nonlinear physics~\cite{Toda1967,Ablowitz1976,Melvin2006,Oxtoby2007,Pelinovsky2007} where the interplay between the nonlinearity and dispersion enables the formation of solitons~-- field configurations maintaining their shape during the propagation. However, to the best of our knowledge, Eq.~\eqref{eq:Lax_QBE_a} and its possible soliton solutions have never been studied before. In such a context, Eq.~\eqref{eq:Lax_QBE_a} is quite exotic as its right-hand side lacks the linear part and only includes nonlinear terms. 


Nonlinear equations~\eqref{eq:Lax_QBE_a} encompass a variety of solutions depending on the chosen initial conditions. As an illustrative example, we investigate the transport of initially localized excitation. Drawing on the intuition from numerically found optimal control for the finite intermediate-scale lattices~\cite{Stepanenko2025}, we anticipate that the fastest way to transport an initially localized excitation in such lattice is in the form of a localized wave packet maintaining its shape during the propagation, i.e. soliton-like solution. Therefore, we aim to compute such coupling controls $J_{m,m+1}(t)$ which transport the localized wavepacket $\ket{\psi(0)}\equiv \ket{\psi_0}$ by one lattice site preserving its spatial distribution $|\psi_m(\tau)|^2=|\psi_{m-1}(0)|^2$ for any $m$ and reaching the fastest possible speed $1/\tau$ of the transfer~-- quantum speed limit~\cite{Deffner2017,ZhenWang2025}.

{\it Results.~}We now proceed to study the soliton-like solutions of Eq.~\eqref{eq:Lax_QBE_a}. In the finite geometry, initial and boundary conditions for Eq.~\eqref{eq:Lax_QBE_a} are defined by the known initial and target quantum states via Eq.~\eqref{eq:WaveFunc}, and the resulting boundary-value problem is solved by the shooting method or gradient descent. \blue{Reduction in the number of variables allows us to treat truly large-scale systems including up to 10 000 qubits, see numerical results in Sec.~VII, Ref.~\cite{Supplement}. This provides two orders of magnitude improvement in terms of size compared to the other systems previously studied in quantum optimal control~\cite{Murphy2010,Orozco2024,Stepanenko2025}.} 

This approach \blue{requires modification} for the effectively infinite lattice, which can be modeled as an array of $N$ qubits closed into ring, Fig.~\ref{fig:1}. Contrary to the finite case, we do not know the initial state $\ket{\psi_0}$ in advance, but rather seek such $\ket{\psi_0}$ which maintains its shape after a round-trip. To ensure that, suitable boundary conditions have to be imposed. Our analysis suggests that soliton-like \blue{moving} solutions  appear if the boundary conditions are chosen antiperiodic, i.e.
\begin{equation}\label{eq:Antiperiodic}
a_{N+1}=-a_1,\mspace{10mu}    \blue{a_{0}=-a_N.}
\end{equation}
%
To expedite numerical analysis, we separate real and imaginary parts of the complex amplitudes $a_m=x_m + i y_m$ in the complex-valued Eq.~\eqref{eq:Lax_QBE_a}:
\begin{eqnarray}
\label{eq:x}
    \partial_t x_m/(2L_0) &=& (x_{m-1}^2 + x_{m+1}^2)y_m \nonumber\\&&- (x_{m-1}y_{m-1}+x_{m+1}y_{m+1})x_m,\\
    \label{eq:y}
    \partial_t y_m/(2L_0) &=& -(y_{m-1}^2 + y_{m+1}^2)x_m \nonumber\\&&+ (x_{m-1}y_{m-1}+x_{m+1}y_{m+1})y_m\:.
\end{eqnarray}
In terms of $x_m$ and $y_m$, $\psi^Q_m=x_m\,\sqrt{2}$, while coupling
\blue{
\begin{equation}\label{eq:Coupling}
J_{m,m+1}=2L_0\,(x_my_{m+1}-x_{m+1}y_m)\:.    
\end{equation}}
\blue{Once we choose $\tau=1$, the value of $L_0$ is uniquely fixed.} Another important observation is that the soliton-like solution corresponds to the localized distribution of $x_m$ amplitudes. However, the amplitudes $y_m$ do not need to be localized.

We seek the numerical solution by the shooting method assuming that the initial distribution of $x_m(0)$ amplitudes is symmetric, i.e. $x_m(0) = x_{N-m+1}(0)$. At the same time, orthogonality of $\ket{x}$ and $\ket{y}$ vectors ensures that the distribution of $y_m(0)$ is antisymmetric $y_m(0) = -y_{N-m+1}(0)$. We request that after time $\tau$ the wavepacket preserves its original shape, i.e. $x_{m+1}(\tau) = x_m(0)$ and $y_{m+1}(\tau) = y_m(0)$. However, due to the twisted boundary conditions, $x_1(\tau)= -x_N(0)$ and $y_1(\tau)= -y_N(0)$.




Additionally, we request that $L_0$ is constant in time, while the real vectors $\ket{x}$ and $\ket{y}$ are normalized by the conditions $\<x|x\>=\<y|y\>=1/2$ and $\<x|y\>=0$ which follow directly from $\<a|a\>=1$ and $\<a|a^*\>=0$. Even though the above conditions make the system overdetermined, we are able to construct a unique solution resembling that in a finite-size array of qubits.



\begin{figure}
    \centering
    \includegraphics[width=1\linewidth]{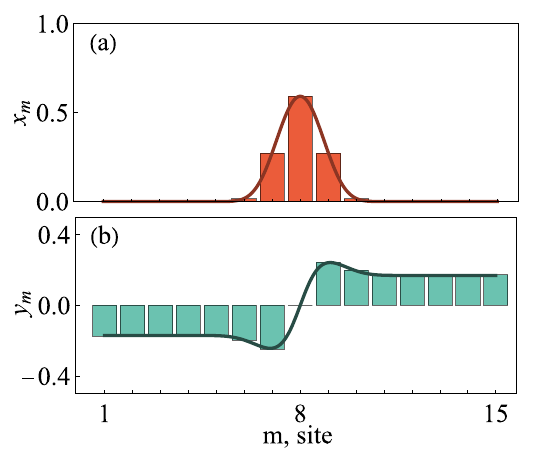}
    \caption{Initial distribution of real $\ket{x(0)}$ (a) and imaginary $\ket{y(0)}$ (b) parts of the Lax eigenvector $\ket{a(0)}$ corresponding to the soliton-like solution in an array of 15 qubits with anti-periodic boundary conditions.}
    \label{fig:2}
\end{figure}

The calculated initial distribution of $x_m$ and $y_m$ amplitudes is shown in Fig.~\ref{fig:2}. As expected, $\ket{x}$ vector related to the wave function features strong localization and occupies just several sites. At the same time, $\ket{y}$ is delocalized, possesses kink-type structure and features non-decaying antisymmetric tails which motivate our choice of anti-periodic boundary conditions. Both histograms Fig.~\ref{fig:2}(a,b) change during the evolution. However, the envelope function shown in Fig.~\ref{fig:2} by the solid lines remains unchanged and simply shifts in space~\cite{Supplement} (Sec.~IV).

\begin{figure}
    \centering
    \includegraphics[width=1\linewidth]{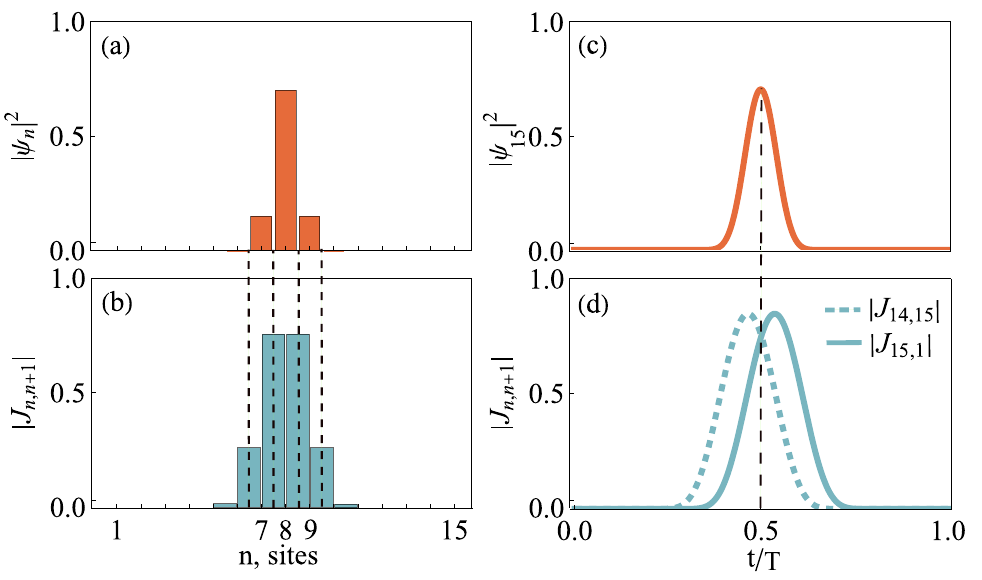}
    \caption{Initial profile of the localized soliton-like solution $|\psi_n(0)|^2$ (a) and associated coupling amplitudes $|J_{n,n+1} (0)|$ (b) in an array of 15 qubits with anti-periodic boundary conditions. Time dependence of the probability $|\psi_{15}|^2$ (c) and the coupling amplitudes $|J_{14,15}|$ (dashed line) and $|J_{15,1}|$ (solid line) (d) over the period $T=N\tau$.}
    \label{fig:3}
\end{figure}

Next, we compare the calculated probability distribution $|\psi_m|^2$ to the pattern of the couplings $J_{m,m+1}$ [Fig.~\ref{fig:3}(a,b)]. The distributions are perfectly correlated: the maximal site population arises exactly between the couplings with the maximal amplitude.
Such behavior is easy to understand. Since the overall coupling resource $J_0$ is limited, it is logical to activate only those coupling links that contribute to the transfer at a given moment of time. The temporal dependence of the probabilities $|\psi_m|^2$ and couplings $J_{m,m+1}$ is described by the envelope functions Fig.~\ref{fig:3}(c,d) which shift with time preserving their shape. This once again points towards the solution in the form of a traveling wave. 

The key characteristic of our solution is the propagation speed. By setting $\tau=1$, we numerically evaluate $J_0=\sqrt{\sum_m J_{m,m+1}^2}$ constant which is equal $J_0=1.13031$. \blue{While this result is obtained for the 15-site lattice closed into ring, the value of $J_0$ does not change if the size of the ring is increased up to 10~000~\cite{Supplement}, Sec.~VII.} This defines the quantum speed limit in our nearest-neighbor-coupled system \blue{under the quadratic constraint on the couplings}.



By inspecting longer evolution times, we observe that the profile of the solution persists even as it makes a round-trip in the ring. However, the wave function acquires an additional quantized geometric phase equal to $\pi$~\cite{Supplement} (Sec.~V). Such quantization is even more intriguing given that the evolution is markedly non-adiabatic, so that at each moment of time the wave function does not correspond to a single dominant instantaneous eigenstate~\cite{Berry2009,WangeSong2025} but 
is rather presented as their superposition~\cite{Supplement}, Sec.~V.

{\it Summary and outlook.} In summary, we have put forward an efficient method to find a time-optimal evolution of a quantum system. \blue{Differently from such established optimal control techniques as GRAPE~\cite{Khaneja2005}, CRAB~\cite{Doria2011} or Krotov~\cite{Goerz2019}, our method not only ensures maximal fidelity of the transport, but is specifically tailored to seek the fastest route, providing a constructive way to achieve the quantum speed limit in various physical settings, as further illustrated in Sec.~IX, Ref.~\cite{Supplement}.}

Combining the Lax pair approach with the quantum brachistochrone technique, we have revealed an integrable nature of time-optimal control problem and substantially reduced the number of variables from approximately $N^2/2$ to $N$, where $N$ is the size of a Hamiltonian. The introduced \blue{Lax eigenvectors} fully capture the dynamics of the system, encoding both the wave function and the Hamiltonian, \blue{whose temporal evolution is synchronized}.

As an illustration of our approach, we have calculated a time-optimal strategy to transfer a single-particle excitation in \blue{a nearest-neighbor-coupled qubit array with the time-varying couplings and constrained sum of their squares. We solved this problem for the lattice with 10~000 qubits both with open and periodic boundary conditions. This exceeds the scale of all other systems previously studied in such context by two orders of magnitude, providing an important advance in the theory of time-optimal quantum evolution.}

\blue{It should be stressed that the developed method is general and applies to various quantum systems. In particular, the End Matter presents the time-optimal strategy to transfer a single-particle excitation in a one-dimensional lattice, where the couplings are constrained to be constant in time. In that case, the fastest strategy is recovered analytically with a rigorous proof of its global optimality.}

\blue{We anticipate that the same method covers few-particle excitations, other types of constraints on the Hamiltonian, 2D or higher-dimensional structures, and, with a suitable generalization, dissipative quantum systems.}


Overall, this study finds a subtle balance between the quantum speed limit and the physical constraints on the Hamiltonian  opening a door to the efficient simulation of large-scale quantum systems and time-optimal quantum computation algorithms.

{\it Acknowledgments.~} We acknowledge Alexander Mikhalychev, Sergei Kilin, Alexey Gorlach, Stanislav Baturin, Igor Shenderovich and Alexey Yulin for valuable discussions. Theoretical models were supported by Priority 2030 Federal Academic Leadership Program. Numerical simulations were supported by the Russian Science Foundation, grant No.~25-42-10012.



 \bibliography{ref1}

\appendix
\clearpage 
\onecolumngrid
\begin{center}
    {\bf End Matter}
\end{center}
\twocolumngrid

\begin{figure}[b!]
    \centering
    \includegraphics[width=0.9\linewidth]{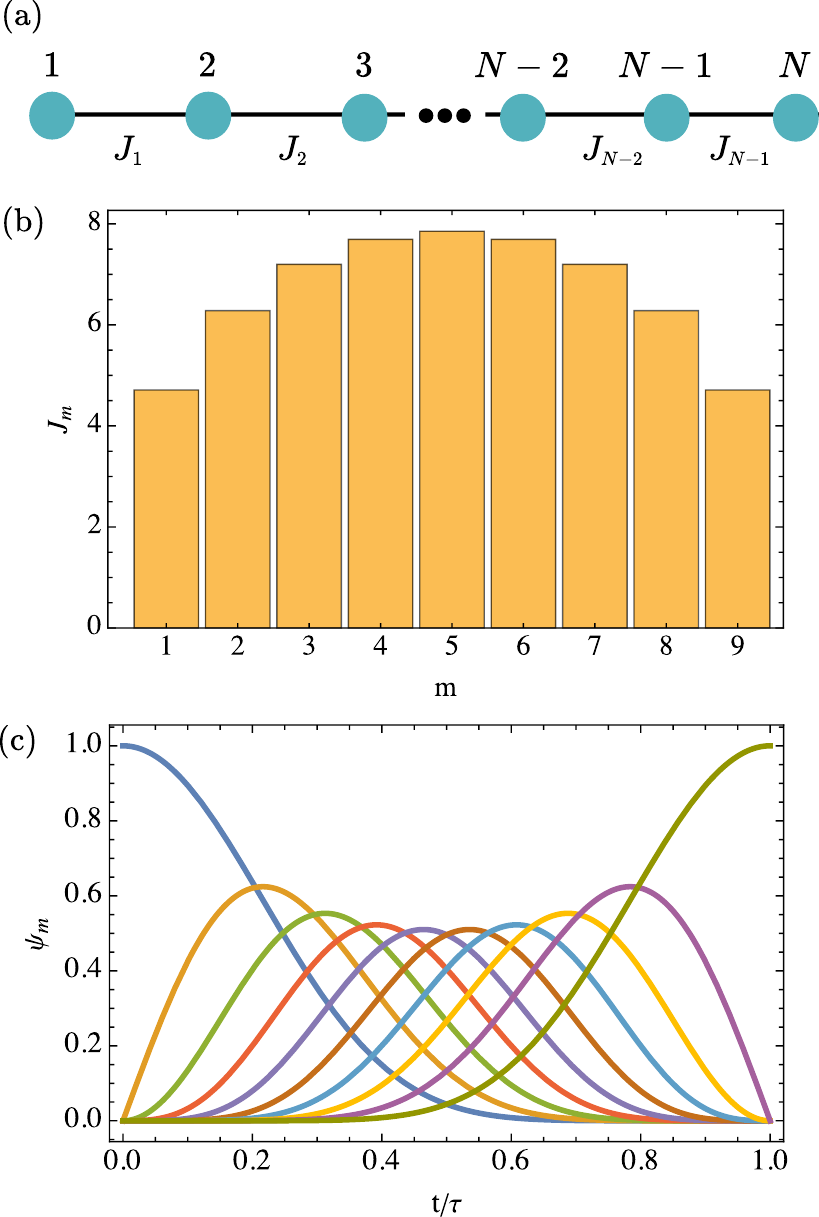}
    \caption{\blue{(a) Scheme of the system with time-independent nearest-neighbor couplings $J_m$ and open boundary conditions. (b) A set of couplings $J_m = \pi\sqrt{m(N-m)}/2$ ensuring perfect transfer for the lattice of $N = 10$ qubits. (c) Evolution of the wave function $\psi_m(t)$ during perfect transfer.}}
    \label{fig:sketch_cc}
\end{figure}

\blue{{\it Time-optimal protocol for the static lattice.} To show the generality of our method, in this section we solve the problem of time-optimal transfer in a different physical system: a lattice of nearest-neighbor-coupled qubits with identical frequencies and time-independent couplings $J_m$, Fig.~\ref{fig:sketch_cc}(a). As previously, the sum of squares of the couplings $J_0^2$ is fixed. Our goal is to find the distribution of time-independent couplings $J_m$ that enables the fastest possible single-particle transfer between the initial state $\ket{1}$ and final state $\ket{N}$.
}


\blue{To ensure that the couplings are time-independent, the cost functional Eq.~\eqref{eq:Cost} should be supplemented by the additional term}
%
\blue{
\begin{equation}
    S_\mu = \sqrt{2}\int\limits_0^\tau  \sum\limits_{m=1}^{N-1} \mu_m \left[\text{Tr}\left(\hat{A}_{m,m+1}\hat{H}\right)-\sqrt{2}J_m\right] dt,
\end{equation}
where $\mu_m$ are additional time-dependent Lagrange multipliers. Variation of the full action $S_{\text{tot}}=S+S_\mu$ with respect to the Lagrange multipliers $\lambda_k$, $\mu_m$ and $\hat{R}$ yields the imposed constraints $\text{Tr}(\hat{H}\hat{D})=0$, $\text{Tr}(\hat{H}\hat{A}_{m,m+1})=\sqrt{2}\,J_m$ and the evolution of the density matrix
\begin{equation}\label{eq:DensityMatrixApp}
\partial_t\hat{\rho} =-i[\hat{H},\hat{\rho}]\:.   
\end{equation}
Variation with respect to the dynamical variables, density matrix $\hat{\rho}$ and Hamiltonian $\hat{H}$, yields
\begin{eqnarray}
&& \partial_t\hat{R} = -i[\hat{H},\hat{R}]\;,\label{eq:Requation}\\
&& \hat{L} = \hat{H}+\hat{D}+\sqrt{2}\sum\limits_{m=1}^{N-1}\mu_m\hat{A}_{m,m+1}\:,\label{eq:LaxDef}
\end{eqnarray}
where we defined the Lax operator as earlier, $\hat{L}=i[\hat{R},\hat{\rho}]$. This definition leads to the same algebraic structure of the Lax operator, Eq.~\eqref{eq:Lax}, while Eqs.~\eqref{eq:DensityMatrixApp},\eqref{eq:Requation} ensure that the operator $\hat{L}$ evolves according to the same Eq.~\eqref{eq:QBE}. The only change is the specific form Eq.~\eqref{eq:LaxDef} of the Lax operator, which now includes the coefficient $\sqrt{2}\,(J_m+\mu_m)$ in front of $\hat{A}_{m,m+1}$ matrix.
}

\blue{
Finally, the couplings $J_m$ are time-independent parameters of the solution, and their choice should extremize the cost functional via $\partial S_{\text{tot}}/\partial J_m=0$, which yields the condition
\begin{equation}
    \int_0^1 \mu_m(t) dt = 0\;.
\end{equation}
Next we perform the same unitary transformation $\hat{Q}=\sum_m i^{m-1}\,\ket{m}\bra{m}$. This reduces the description to the single Lax eigenvector $\ket{a} \equiv \hat{Q}\ket{\alpha}=\ket{x}+i\ket{y}$ with the Lax eigenvalue $L_0$, while the second eigenvector is obtained by complex conjugation due to chiral symmetry. In this representation, the governing equations read
%
%
\begin{eqnarray}
     \partial_t x_m = J_{m-1} x_{m-1} - J_m x_{m+1}\;,\label{eq:Schr1}\\
     \partial_t y_m = J_{m-1} y_{m-1} - J_m y_{m+1}\;,\label{eq:Schr2}\\
     \label{eq:integral}
     \int_0^\tau (x_m y_{m+1}-x_{m+1}y_m)\,dt = J_m\,\tau/(2 L_0)
\end{eqnarray}
with the boundary conditions
\begin{equation}\label{eq:Boundary1}
x_m(0) = \delta_{1,m}/\sqrt{2},\mspace{6mu} x_m(\tau) = \pm\delta_{N,m}/\sqrt{2}\:. 
\end{equation}
Having a solution $x_m(t),y_m(t),J_m$ to the above equations, one can construct another one with the same transfer time $\tau$:
\begin{gather}
\tilde{x}_{m}(t)= \pm x_{N+1-m}(\tau-t)\:,\label{eq:Trans1}\\
\tilde{y}_{m}(t)=\mp y_{N+1-m}(\tau-t)\:,\label{eq:Trans2}\\
\tilde{J}_m=J_{N-m}\:,\label{eq:Trans3}
\end{gather}
where the choice of the sign is dictated by the boundary condition  for $x_m(\tau)$, Eq.~\eqref{eq:Boundary1}. As the optimal solution corresponds to the local extremum of the cost $S_{\text{tot}}$, it is non-degenerate. Thus, the transformation Eqs.~\eqref{eq:Trans1}-\eqref{eq:Trans3} yields the same solution and therefore the pattern of the couplings is necessarily symmetric:
\begin{equation}\label{eq:CouplingSym}
J_m=J_{N-m}\:.
\end{equation}
Such reflection symmetry of the Hamiltonian Eq.~\eqref{eq:CouplingSym} guarantees that the evolution of the state in the interval $\tau<t<2\tau$ mirrors its evolution in the earlier period $0<t<\tau$ and hence after time $2\tau$ the excitation must return to the initial quantum state:
\begin{equation}\label{eq:PeriodicityCond}
    \ket{\psi(2\tau)}=e^{i\theta}\,\ket{\psi(0)}\:.
\end{equation}
Since the wave function in this representation is real, the geometric phase $\theta$ is either $0$ or $\pi$.
}

\blue{
On the other hand, since the Hamiltonian is time-independent, the wave function can be expanded in the basis of the stationary states $\ket{k_m}$ with the energies $E_m$ yielding
\begin{equation}
\ket{\psi(t)}=\sum\limits_{m=1}^{N}\,c_m\,e^{-iE_m t}\,\ket{k_m}\:,
\end{equation}
where energies $E_m$ are ordered in ascending order. The periodicity condition Eq.~\eqref{eq:PeriodicityCond} then translates into 
\begin{equation}\label{eq:Spectrum}
E_m=\frac{2\pi n_m-\theta}{2\tau}\:,
\end{equation}
where $n_m$ is an integer. Since the spectrum is symmetric with respect to zero energy due to chiral symmetry, $\theta=\pi$ for even $N$ and $\theta=0$ for odd $N$. Equation~\eqref{eq:Spectrum} strongly restricts possible couplings $J_m$ requiring that the energies in the spectrum of the time-independent lattice are related as integer numbers. Furthermore, the transfer can only happen if all couplings are nonzero, i.e. the lattice is connected. Then the tri-diagonal structure of the Hamiltonian guarantees that the spectrum is non-degenerate, i.e. $n_m$ with the different $m$ cannot coincide.}

\blue{
To choose the fastest transfer strategy, we present the Hamiltonian as $\hat{H}=\sum_m\,E_m\,\ket{k_m}\bra{k_m}$ and evaluate the coupling resource
\begin{equation}
J_0^2=\frac{1}{2}\,\text{Tr}\,\hat{H}^2=\frac{1}{2}\,\sum_m E_m^2=\frac{1}{8\tau^2}\,\sum_m (2\pi n_m-\theta)^2\:.
\end{equation}
As we show~\cite{Supplement}, Sec.~VIII, the eigenvalues are bounded $|E_m|\geq \pi |2m-N-1|/(2\tau)$ and hence
\begin{equation}\label{eq:Bound}
    J_0\tau \geq \pi\sqrt{N(N^2-1)/24}\;.
\end{equation}
We observe that the smallest possible value of $J_0\,\tau$ is achieved when $n_m$ are chosen as consecutive integer numbers ($n_{m+1}-n_{m}=1$), and any other choice further increases $J_0\,\tau$.
}

\blue{
Equidistant symmetric energy spectrum, tridiagonal structure of the Hamiltonian and the boundary conditions Eq.~\eqref{eq:Boundary1} define a unique transfer strategy. It appears, that such strategy is already known in the literature under the name of {\it perfect transfer}~\cite{Christandl2004}. Our analysis proves that the perfect transfer is the fastest possible way to deliver a single-particle excitation from one edge of the tight-binding lattice of coupled identical qubits to another, provided the sum of squares of the couplings is fixed to $J_0^2$. This finding correlates with the earlier work~\cite{Yung2006}, which studied the optimality of perfect transfer under different constraint, fixed maximal coupling in the lattice.
}

\blue{
For consistency, we briefly summarize the optimal solution below. The optimal coupling distribution
\begin{equation}
J_m=\frac{B}{2}\,\sqrt{m\,(N-m)}\:,
\end{equation}
where $B$ is constant and $m=1,2,\dots (N-1)$ is illustrated in Fig.~\ref{fig:sketch_cc}(b), featuring the maximal couplings in the middle of the lattice. The energy spectrum is equidistant $E_k=B\,k$, where $k=-s,-s+1,\dots,s-1,s$ and $s=(N-1)/2$, resembling that for the spin in magnetic field. The evolution of the wave function is solved analytically and reads
\begin{equation}
\label{eq:psi_cc}
\psi_m(t)=\sqrt{C_{N-1}^{m-1}}\,\cos^{N-m}\left(\frac{Bt}{2}\right)\,\sin^{m-1}\left(\frac{Bt}{2}\right)\:,
\end{equation}
where $C_{N}^k=\frac{N!}{k!\,(N-k)!}$ is the binomial coefficient, so that at time $\tau=\pi/B$ the entire population is transferred to the last qubit of the lattice. The evolution of the probability amplitudes is depicted in Fig.~\ref{fig:sketch_cc}(c). Finally, the coupling resource Eq.~\eqref{eq:Bound}
scales as $N^{3/2}$, which means substantially slower transport compared to the scenario with switchable couplings.
}

\blue{
Note also that the expansion for the auxiliary vector $\ket{y}$ in terms of the stationary states $\ket{k_m}$ can be found from Eq.~\eqref{eq:integral}, which yields the linear system of equations with respect to the unknown superposition coefficients.
}

\clearpage 
\onecolumngrid
\section*{Supplementary Materials} 
\setcounter{section}{0} 
\setcounter{equation}{0} 
\setcounter{page}{1}
\renewcommand{\theequation}{S\arabic{equation}} 
\renewcommand{\thefigure}{S\arabic{figure}} 

\section{I. Derivation of quantum brachistochrone equation}

In this section, we derive quantum brachistochrone equation together with the boundary conditions. First, we note that the time variable in the Schr{\"o}dinger equation can be rescaled via $t\rightarrow g(t)$, where $g(t)$ is an arbitrary monotonically increasing function. Such rescaling renormalizes the Hamiltonian via
\begin{equation}
\hat{H}\rightarrow \hat{H}/\partial_t g\:.
\end{equation}
Therefore, if the norm of the original Hamiltonian is constrained $\|\hat{H}\| = \sqrt{\text{Tr}\,\hat{H}^2}\leq \sqrt{2}\,J_0$ reaching the limit $\sqrt{2}\,J_0$ only at some particular moments of time, after the transformation it can be made maximal and constant $\|\hat{H}\| = \sqrt{2}\,J_0$ during the entire evolution, which further decreases the evolution time. Therefore, from now on we study the Hamiltonians with the constant norm $\|\hat{H}\|$.

As discussed in the main text, minimization of the transfer time $\tau$ for the fixed norm of the Hamiltonian $\|\hat{H}\|$, prescribed physical constraints and boundary conditions is equivalent to minimizing the norm of the Hamiltonian $\|\hat{H}\|$ for the fixed transfer time $\tau$ with the same constraints and conditions. This dictates the cost functional
%
\begin{equation}\label{eqs:Cost1}    \ell=\int\limits_0^\tau\,\|\hat{H}\|\,dt+\int\limits_0^\tau\,\text{Tr}(\hat{H}\hat{D}')\,dt+
\int\limits_0^\tau\, \mathrm{Tr}\left(\hat{R}_t'(\partial_t\hat{\rho}+i[\hat{H},\hat{\rho}])\right)\;dt.
\end{equation}

However, fixed norm $\|\hat{H}\|$ of the Hamiltonian allows us to simplify the calculation and replace the functional Eq.~\eqref{eqs:Cost1} by a simpler construction
%
\begin{equation}\label{eqs:Cost2}    
S=\frac{1}{2}\int\limits_0^\tau\,\|\hat{H}\|^2\,dt+\int\limits_0^\tau\,\text{Tr}(\hat{H}\hat{D})\,dt
\int\limits_0^\tau\, \mathrm{Tr}\left(\hat{R}_t(\partial_t\hat{\rho}+i[\hat{H},\hat{\rho}])\right)\;dt.
\end{equation}
Indeed, the condition $\delta S=0$ translates into $\delta l=0$ with the identification $\hat{D}'=\hat{D}/\|\hat{H}\|$ and $\hat{R}_t'=\hat{R}_t/\|\hat{H}\|$. We therefore work with the functional Eq.~\eqref{eqs:Cost2}.

The cost functional $S$ depends on the dynamical variables $\hat{H}$, $\hat{\rho}$ and Lagrange multipliers $\hat{R}$, $\lambda_k$, included in $\hat{D} = \sum_k \lambda_k\hat{D}_k$. Computing the variation of $S$ with respect to the dynamical variables $\hat{H}$ and $\hat{\rho}$, we recover:
\begin{eqnarray}
\label{eq:em1}
    \hat{H}+\hat{D}+i[\hat{\rho},\hat{R}] = 0\;,\\
    \label{eq:em2}
    \partial_t\hat{R}+i[\hat{H},\hat{R}] = 0\;.
\end{eqnarray}
Similarly, variation with respect to Lagrange multipliers $\lambda_k$ and $\hat{R}$ yields
\begin{eqnarray}
\label{eq:em3}
    \mathrm{Tr}\left(\hat{H}\hat{D}_k\right) = 0\;,\\
    \label{eq:em4}
    \partial_t\hat{\rho}+i[\hat{H},\hat{\rho}] = 0\;.
\end{eqnarray}
These equations of motion correspond to the extremum condition $\delta S = 0$. Next we define the operator 
\begin{equation}\label{eqS:QB_sol}
\hat{L}=i[\hat{R},\hat{\rho}]\:.    
\end{equation}
Due to Eq.~\eqref{eq:em2}, $\hat{L} = \hat{H} +\hat{D}$. The operators $\hat{R}$ and $\hat{\rho}$ evolve according to Eqs.~\eqref{eq:em2}, 
\eqref{eq:em4}, which yields quantum brachistochrone equation for the operator $\hat{L}$:
\begin{eqnarray}
    \label{eqS:QB}
    \partial_t\hat{L}&=&-i\left[\hat{H},\hat{L}\right]\;.
\end{eqnarray}
with a solution
\begin{equation}
\hat{L}(t)=\hat{U}(t)\,\hat{L}(0)\,\hat{U}^\dagger(t)\:.
\end{equation}

We note that one may consider a unitary evolution operator $\hat{U}$ as a dynamical variable instead of $\hat{H}$ and arrive to the same equations of motion. In this case, since $\delta\hat{U}$ at the boundaries is nonzero, is it more intuitive to consider the cost functional with the boundary terms
\begin{equation}
\label{eqs:Cost2b}    
S=\frac{1}{2}\int\limits_0^\tau\,\|\hat{H}\|^2\,dt+\int\limits_0^\tau\,\text{Tr}(\hat{H}\hat{D})\,dt+\left.\mathrm{Tr}\left(\hat{R}_t(\hat{\rho}_t-\hat{U} \hat{\rho}_0\hat{U}^\dagger)\right)\right|_0^\tau\:.
\end{equation}

\section{II. Lax operator structure}

In this section, we discuss the structure of the Lax operator $\hat{L}$, its eigenvalues and eigenvectors. First, it is instructive to note that Eq.~\eqref{eqS:QB} is a Lax equation~\cite{Babelon2003}. As a consequence, general theory~\cite{Babelon2003} guarantees that the eigenvalues $\gamma$ of the Lax operator do not depend on time.

Below, we reproduce this argument. Assume that $\ket{\alpha(t)}$ is an eigenvector of the Lax operator $\hat{L}(t)$ corresponding to the eigenvalue $\gamma(t)$:
\begin{equation}
 \hat{L}\,\ket{\alpha}=\gamma\,\ket{\alpha}\:.   
\end{equation}
Differentiating this identity with respect to time we recover
\begin{equation}
\partial_t\,\hat{L}\,\ket{\alpha}+\hat{L}\,\ket{\partial_t\,\alpha}=\partial_t \gamma\,\ket{\alpha}+\gamma\,\ket{\partial_t\alpha}\:.
\end{equation}
Next we take into account that the eigenvector $\alpha$ evolves according to the Schr\"odinger equation
\begin{equation}
\ket{\partial_t\alpha}=-i\hat{H}\ket{\alpha}\:,
\end{equation}
while the Lax operator evolves according to Eq.~\eqref{eqS:QB}. This immediately yields
\begin{equation}
    \partial_t\gamma\,\ket{\alpha}=0
\end{equation}
for the arbitrary eigenvector $\ket{\alpha}$. Hence, we conclude that all eigenvalues of the Lax operator are time-independent~\cite{Babelon2003}.

Moreover, Eq.~\eqref{eqS:QB_sol} further constrains the structure of the Lax operator:
\begin{eqnarray}
    \hat{L}=i\hat{R}\hat{\rho}-i\hat{\rho}\hat{R}\;,
\end{eqnarray}
where $\hat{\rho} = \ket{\psi}\bra{\psi}$ and $\hat{R}$ is a Hermitian traceless Lagrange multiplier matrix. We introduce the vector $\ket{r} \equiv \hat{R}\ket{\psi}$. Since $\hat{R}$ is an unknown matrix, $\ket{r}$ may not be orthogonal to $\ket{\psi}$. We use the Gram–Schmidt process to construct a set of two orthogonal vectors:
\begin{eqnarray}
    \ket{e_1} &=& \ket{\psi}\\
    \ket{e_2} &=& (\ket{r} - c\ket{\psi})/L_0
\end{eqnarray}
where $c = \bra{\psi}\hat{R}\ket{\psi}$ and $L_0 = \sqrt{\bra{r}r\rangle-c^2}$.

In this basis, Lax operator takes the following form
\begin{eqnarray}
    \hat{L}=i\ket{r}\bra{\psi}-i\ket{\psi}\bra{r} = iL_0\ket{e_2}\bra{e_1}-iL_0\ket{e_1}\bra{e_2} \;.
\end{eqnarray}

Now it is straightforward to show that the vectors $\ket{\alpha} = (\ket{e_1}+i\ket{e_2})/\sqrt{2}$ and $\ket{\beta} = (\ket{e_1}-i\ket{e_2})/\sqrt{2}$ are the normalized eigenvectors of the Lax operator
\begin{eqnarray}
   & \hat{L}\ket{\alpha}= L_0\,\ket{\alpha}\;,\\
   & \hat{L}\ket{\beta}= -L_0 \ket{\beta}\;,\\
   & \hat{L}= L_0\ket{\alpha}\bra{\alpha}-L_0\ket{\beta}\bra{\beta}\;.
\end{eqnarray}
Thus, we derive
\begin{eqnarray}
    \ket{\alpha} &=& (\ket{\psi}+i(\hat{R}\ket{\psi}-c\ket{\psi})/L_0)/\sqrt{2}\;,\\
    \ket{\beta}&=&(\ket{\psi}-i(\hat{R}\ket{\psi}-c\ket{\psi})/L_0)/\sqrt{2}\;.
\end{eqnarray}
Note that $\ket{\alpha}+\ket{\beta}=\sqrt{2}\,\ket{\psi}$. 
Moreover, we highlight that all three vectors evolve according to the Schr\"odinger equation and this linear relation holds at the arbitrary moment of time $t$, i.e.
\begin{equation}\label{eqS:wavefunc}
    \ket{\psi(t)}=\frac{1}{\sqrt{2}}\,\left(\ket{\alpha(t)}+\ket{\beta(t)}\right)\,
\end{equation}
which provides the connection between the eigenvectors of the Lax operator and the actual state of the quantum system.

At the same time, since the Lax operator evolves as $\hat{L}(t)=\hat{U}(t)\hat{L}(0)\hat{U}^\dagger(t)$, the analysis above yields
\begin{equation}\label{eqS:Lax}
 \hat{L}(t)=L_0\ket{\alpha(t)}\bra{\alpha(t)}-L_0\ket{\beta(t)}\bra{\beta(t)}\;.  
\end{equation}
On the other hand, $\hat{L}=\hat{H}+\hat{D}$, where $\hat{D}=\sum_p \lambda_p\,\hat{D}_p$. By projecting out known $\hat{D}_p$ matrices, we recover the elements of the Hamiltonian in terms of the Lax operator eigenvectors and the constant factor $L_0$.

Thus, to fully solve quantum optimal control problem, one only needs to determine the temporal dependence of the vectors $\ket{\alpha(t)}$ and $\ket{\beta(t)}$. Moreover, as we show below, in many situations  the two vectors are related to each other in a simple way.

\section{III. Closure of $\mathcal{A}$ subspace}

First, we define the operator $\hat{A}_{m,m+p} \in \mathcal{A}$ as in the main text
\begin{equation}
    \hat{A}_{m,m+p}=\frac{1}{\sqrt{2}}\,(i^{p-1}\ket{m}\bra{m+p}+i^{1-p}\ket{m+p}\bra{m})\:,
\end{equation}
where $\ket{m}=\cra{m}\,\ket{0}$. Note that this matrix satisfies antisymmetry condition $\hat{A}_{m,m+p}=-\hat{A}_{m+p,m}$. To demonstrate $i[\mathcal{A},\mathcal{A}]\subseteq\mathcal{A}$, we compute the following commutator
\begin{eqnarray}
  i\sqrt{2}[\hat{A}_{m,m+p},\hat{A}_{n,n+q}]=\delta_{n,m+p}\,\hat{A}_{m,m+p+q}-\delta_{m+p,n+q}\,\hat{A}_{m,m+p-q}-\delta_{nm} \hat{A}_{m+p,m+q}+\delta_{m,n+q} A_{m+p,m-q}\:. 
\end{eqnarray}
Thus, we recover that the commutator of the two matrices from the $\mathcal{A}$ subspace yields the linear combination of the matrices belonging to the same subspace.

Next we verify that the commutator of the two matrices $A_k\in\mathcal{A}$ and $B_l\in\mathcal{B}$ belongs to the $\mathcal{B}$ subspace, where $\mathcal{B}$ complements $\mathcal{A}$ to the complete space $\mathcal{M}$ of traceless Hermitian matrices. To that end, we compute the projection of the commutator $i[A_k,B_l]$ onto the basis matrix $A_m\in\mathcal{A}$:
\begin{equation}
\zeta_m=i\,\text{Tr}\left(A_m[A_k,B_l]\right)=i\,\text{Tr}\left(A_mA_kB_l-A_m B_l A_k\right)=i\,\text{Tr}\left(A_mA_kB_l-A_kA_m B_l\right)=i\,\text{Tr}\left([A_m,A_k]B_l\right)\:.
\end{equation}
However, as shown above, the commutator $i[A_m,A_k]$ must belong to the $\mathcal{A}$ subspace, while $\hat{B}_l\in\mathcal{B}$. Hence, the projection $\zeta_m=0$ which proves that the commutator $i[\mathcal{A},\mathcal{B}]\in\mathcal{B}$. 

Following these algebraic properties, we separate the Lax operator into two parts, $\hat{L}^A\in\mathcal{A}$ and $\hat{L}^B\in\mathcal{B}$. Given that $\hat{H}\in\mathcal{A}$, Eq.~\eqref{eqS:QB} splits into two independent parts:
\begin{equation}\label{eq:QBE-reduced}
\partial_t\hat{L}^A=i[\hat{L}^A,\hat{H}]    
\end{equation}
and $\partial_t\hat{L}^B=i[\hat{L}^B,\hat{H}]$. As a result, the dynamics of Lagrange multipliers from the $\mathcal{B}$ subspace get fully decoupled, and we only need to solve Eq.~\eqref{eq:QBE-reduced}.

Furthermore, if we are interested only in the single-particle excitations, the wave function spans just $N$-dimensional single-particle sector of the Hilbert space. In such case, the operator $\hat{A}_{m,m+p}$ is a traceless Hermitian $N\times N$ matrix
\begin{eqnarray}
    \hat{A}_{m,m+p} &=& (i^{p-1}\hat{E}_{m,m+p}+i^{1-p}\hat{E}_{m+p,m})/\sqrt{2}\:,
\end{eqnarray}
where $1\leq m\leq N-1$ and $1\leq p\leq N-m$ and the matrix $E_{mn}$ has the elements $(E_{mn})_{pq}=\delta_{mp}\delta_{nq}$. The elements of the complementary space $\mathcal{B}$ are defined as
\begin{eqnarray}
    \hat{B}_{m,m+p} &=& (i^{p}\hat{E}_{m,m+p}+i^{-p}\hat{E}_{m+p,m})/\sqrt{2}\:,\\
    \hat{B}_{m,m} &=&\left(\sum_{k=1}^m\hat{E}_{kk}-m\hat{E}_{mm}\right)\sqrt{2/(m^2+m)}\:,
\end{eqnarray}
where $1\leq m< N$ and $1\leq p\leq N-m$. The Hamiltonian and the Lax operator take the form
\begin{eqnarray}
    \hat{H} &=& \sqrt{2}\sum\limits_{m=1}^{N-1}\,J_{m,m+1}\hat{A}_{m,m+1}+\sqrt{2}J_{1,N}\hat{A}_{1,N}\:,\\
    \hat{L}^A &=& \sqrt{2}\sum\limits_{m=1}^{N-1}\sum\limits_{p=2}^{N-m}\lambda_{m,m+p}^A\hat{A}_{m,m+p}-\sqrt{2}\lambda_{1,N}^A\hat{A}_{1,N}+\hat{H}\:,\\
    \hat{L}^B &=& \sqrt{2}\sum\limits_{m=1}^{N-1}\sum\limits_{p=0}^{N-m}\lambda_{m,m+p}^B\hat{B}_{m,m+p}\:.
\end{eqnarray}
%

Importantly, both the Hamiltonian and the $\mathcal{A}$-projected Lax operator $\hat{L}^A$ possess {\it chiral symmetry}. To demonstrate that, we introduce chiral symmetry operator 
\begin{equation}
  \hat{\sigma} = \sum_m (-1)^{m-1}\ket{m}\bra{m}\,\hat{K}\:,  
\end{equation}
where $\hat{K}$ denotes complex conjugation. As it is straightforward to verify, an anticommutator $\left\{\hat{A}_{m,m+p},\hat{\sigma}\right\}=0$, which ensures that
\begin{gather}
    \left\{\hat{H},\hat{\sigma}\right\}=0\:,\\
    \left\{\hat{L}^A,\hat{\sigma}\right\}=0\:.
\end{gather}

As a consequence of the chiral symmetry, each eigenvector $\ket{\alpha}$ of the Lax operator $\hat{L}^A$ with an eigenvalue $L_0$ is accompanied by another eigenvector $\ket{\beta}=\hat{\sigma}\ket{\alpha}$ with the opposite eigenvalue $-L_0$. Thus, chiral symmetry ensures that the spectrum of the Lax operator is symmetric.

Of course, the same symmetry applies to the spectrum of the Hamiltonian $\hat{H}$ with the same connection between the instantaneous eigenvectors.


To make the treatment more transparent, it is convenient to perform a unitary transformation captured by the diagonal matrix $\hat{Q}$ with the entries $Q_{mm}=i^{m-1}$, i.e. $\hat{Q} =\sum_m i^{m-1}\hat{E}_{m,m}$. The basis matrices are transformed as
\begin{eqnarray}
    \hat{A}_{m,m+p}^Q &=& \hat{Q}\hat{A}_{m,m+p}\hat{Q}^{-1} 
    =  (-i\,\hat{E}_{m,m+p}+i\,\hat{E}_{m+p,m})/\sqrt{2}\:,\\
    \hat{B}_{m,m+p}^Q &=& \hat{Q}\hat{B}_{m,m+p}\hat{Q} ^{-1}
    =  (\hat{E}_{m,m+p}+\hat{E}_{m+p,m})/\sqrt{2}\:,\\
    \hat{B}_{m,m}^Q &=& \hat{Q}\hat{B}_{m,m}\hat{Q}^{-1} = \left(\sum_{k=1}^m\hat{E}_{kk}-m\hat{E}_{mm}\right) \sqrt{2/(m^2+m)}\:.
\end{eqnarray}
Thus, the matrices $\hat{A}_{m,m+p}^Q$ are imaginary, while $\hat{B}_{m,m+p}^Q$ and $\hat{B}_{m,m}^Q$ are  real.


The part of the Lax operator $\hat{L}_Q^A$ contains only the matrices $\hat{A}^Q_{m,m+p}$ which are purely imaginary. These matrices are multiplied by the real coefficients $J_{m,m+1}$ and $\lambda_{m,m+p}$, which renders $\hat{L}_Q^A$ imaginary.

Now we inspect the eigenvector of the Lax operator with a positive eigenvalue: $\hat{L}_Q^A\ket{a} =L_0\ket{a}$. Since $\hat{L}_Q^A$ is imaginary, complex conjugation yields
\begin{equation}
   \hat{L}_Q^A\ket{a^*} =-L_0\ket{a^*}\:.
\end{equation}
This means that the second Lax eigenvector 
\begin{equation}\label{eq:conjugation}
    \ket{b}=\ket{a^*}\:.
\end{equation}
It should be stressed that the connection Eq.~\eqref{eq:conjugation} between the Lax eigenvectors does not depend on the initial state of the quantum system $\ket{\psi(0)}$, being a direct consequence of the chiral symmetry.

\section{IV. Structure of the solution}

\subsection{A. Numerical results}
In this section, we discuss in detail the spatial structure of numerically found soliton-like solution and its temporal evolution. Building on our recent results~\cite{Stepanenko2025} for finite intermediate-scale quantum systems, we consider a localized, center-symmetric wave packet $|\psi_m(0)| = |\psi_{N+1-m}(0)|$. Also, following the notations in the main text, we assume that the transformed wave function $\ket{\psi_Q}=\hat{Q}\ket{\psi}$ is purely real and is given by the expression
\begin{equation}\label{eqs:PsiTransformed}
\ket{\psi_Q} = (\ket{a}+\ket{a^*})/\sqrt{2}\:.    
\end{equation}
The center-symmetric structure of the wave function results in the symmetry of the real part of the eigenvector $\ket{x}=\ket{\psi_Q}/\sqrt{2}$:
\begin{equation}\label{eqs:xsym}
    x_m(0) = x_{N+1-m}(0)\:,
\end{equation}
which suggests a maximum of the wave function in the middle site $p=(N+1)/2$. Note that an alternative option $x_m(0)=-x_{N+1-m}(0)$ does not work, as it predicts zero of the wave function in the same middle site $p$.

The Schr\"odinger equation for the eigenmode $\ket{a}$ presented as a system of differential equations for complex components $a_m$ of the eigenmode
\begin{eqnarray}
\label{eqS:Lax_QBE_a}
    i\partial_t a_m/L_0 &=& (|a_{m-1}|^2 + |a_{m+1}|^2)\,a_m - (a_{m-1}^2+a_{m+1}^2)\,a_m^*\;,
\end{eqnarray}
can be replaced by the system of equations for the real variables $x_m$ and $y_m$: 
\begin{eqnarray}
\label{eqS:x}
    \partial_t x_m/2L_0 &=& (x_{m-1}^2 + x_{m+1}^2)y_m - (x_{m-1}y_{m-1}+x_{m+1}y_{m+1})x_m\:,\\
    \label{eqS:y}
    \partial_t y_m/2L_0 &=& -(y_{m-1}^2 + y_{m+1}^2)x_m + (x_{m-1}y_{m-1}+x_{m+1}y_{m+1})y_m\:.
\end{eqnarray}

Next we examine the symmetry of $y_m$ distribution. To that end, we recall that the Lax eigenvector is presented as $\ket{a}=\ket{x}+i\ket{y}$ with the real vectors $\ket{x}$ and $\ket{y}$ and satisfies the properties $\<a|a\>=1$, $\<a^*|a\>=0$. This translates into the conditions
\begin{eqnarray}
    &&\<x|x\>=\<y|y\>=1/2\:,\\
    && \<x|y\>=0\:.\label{eqs:ScalarProd}
\end{eqnarray}
Moreover, as dictated by the inversion symmetry of the problem, $\ket{x}$ and $\ket{y}$ must be either even or odd under the inversion transformation $a_m\rightarrow a_{N+1-m}$. Equation~\eqref{eqs:ScalarProd} then suggests that $y_m$ is antisymmetric:
\begin{equation}\label{eqs:ysym}
y_{m}(0) = -y_{N+1-m}(0)\:,    
\end{equation}
and as a consequence $y_p(0)=0$. We use Eqs.~\eqref{eqs:xsym}, \eqref{eqs:ysym} as the initial conditions for our numerical procedure.

To find the vectors $\ket{x(0)}$ and $\ket{y(0)}$ we use the shooting method. Equations~\eqref{eqS:x}-\eqref{eqS:y} have $(2N+1)$ unknown variables $x_m$, $y_m$ and $L_0$, while we have only $2N$ differential equations. The value of $L_0$ is unknown, but it does not depend on time and thus $\partial_tL_0 = 0$.

\begin{figure}
    \centering
    \includegraphics[width=\linewidth]{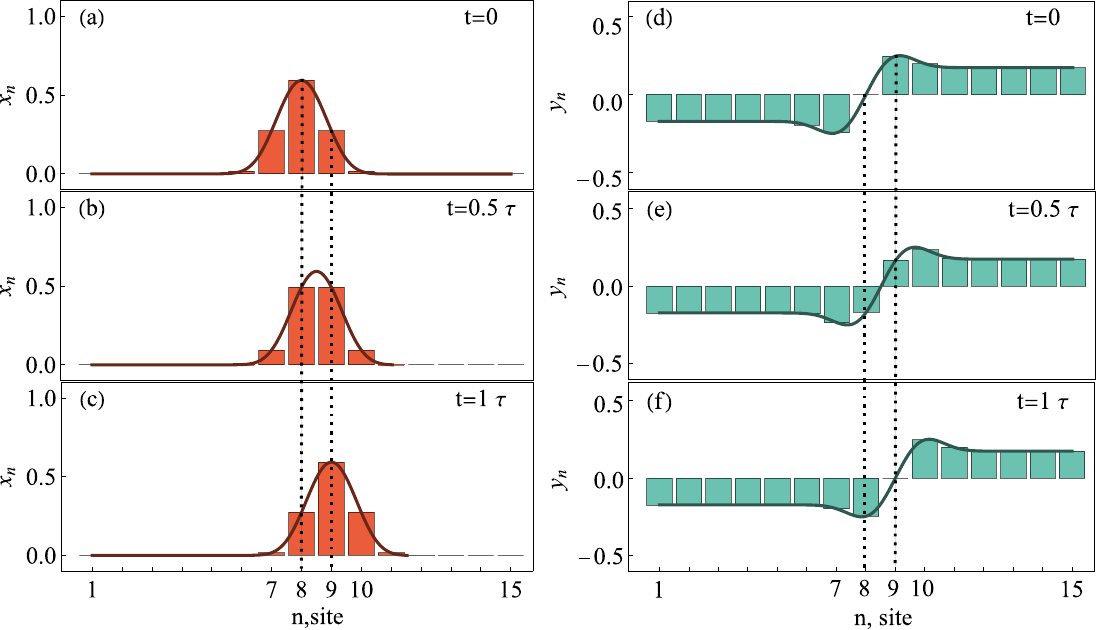}
    \caption{The structure of the Lax eigenmode $\ket{a}$ for the $N=15$ qubit array closed into ring with anti-periodic boundary conditions. Real part $\ket{x}$ at time moments (a) $t=0$, (b) $t=\tau/2$, (c) $t=\tau$. Imaginary part $\ket{y}$ at times (a) $t=0$, (b) $t=\tau/2$, (c) $t=\tau$. The dark solid line shows the time evolution for the 15-th qubit over time $N\tau$, rescaled via $n = Nt/\tau$ and shifted to match the positions of the maxima.}
    \label{fig:S1}
\end{figure}

To find a solution to the system of $(2N+1)$ differential equations, we need $(2N+1)$ initial and boundary conditions. Using the symmetry of $\ket{x(0)}$ and $\ket{y(0)}$ vectors at the initial moment of time, we recover $N$ initial conditions. In addition, we require that the wave packet retains its shape after time $\tau$ when the center of mass shifts by one lattice site: $\ket{x(\tau)}=\hat{T}\ket{x(0)}$ and $\ket{y(\tau)}=\hat{T}\ket{y(0)}$, where $\hat{T} = \sum_m\ket{m+1}\bra{m}$ is a right-translating operator. This yields $2\,N$ boundary conditions and makes the overall system overdetermined. However, as we show, the solution still exists.

For the small arrays with $N<5$ closed into ring we used almost random  initial guess, respecting the symmetry of $x_m$ and $y_m$. The shooting method converges, which provides us an intuition about the structure of the solution. This allows us to construct a relatively good initial guess for the larger arrays.

The numerical results for the vectors $\ket{x}$ and $\ket{y}$ are illustrated in Fig.~\ref{fig:S1}. Interestingly, the curve describing the evolution of the wave function at a single site $m=15$ during the time $T=N\,\tau$ almost perfectly fits the shape of the spatial distributions at all moments of time with the absolute error below $10^{-5}$. This suggests that the amplitudes $x_m=x(m-t/\tau)$, $y_m=y(m-t/\tau)$, and the entire wave packet has a structure of the propagating wave. To illustrate that, we show the spatial distribution at the initial $t=0$ [Fig.~\ref{fig:S1}(a,d)], intermediate $t=\tau/2$ [Fig.~\ref{fig:S1}(b,e)], and final $t = \tau$ [Fig.~\ref{fig:S1}(c,f)] moments of time.

\begin{figure}
    \centering
    \includegraphics[width=\linewidth]{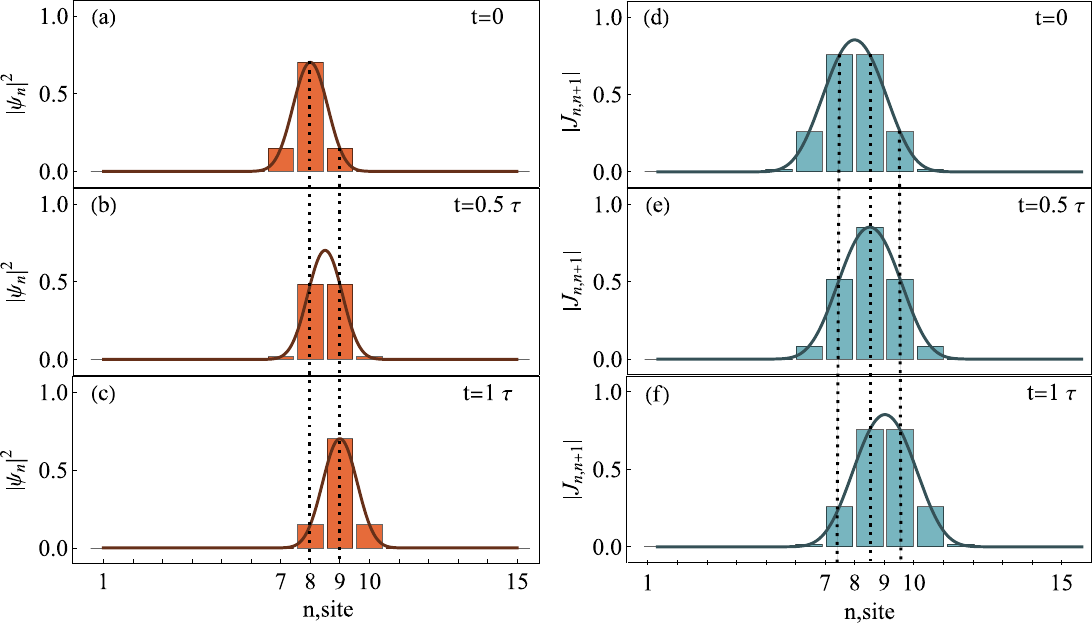}
    \caption{The structure of the site occupation $|\psi_m|^2$ for the 15-th qubit in the array with anti-periodic boundary conditions at times (a) $t=0$, (b) $t=\tau/2$, (c) $t=\tau$. Spatial distribution of the coupling amplitudes $J_{m,m+1}$ at times (a) $t=0$, (b) $t=\tau/2$, (c) $t=\tau$. Dark solid line shows the time evolution of the 15-th qubit population and 15-th coupling amplitude over time $N\tau$, rescaled via $n = Nt/\tau$ and shifted to match the positions of the maxima.}
    \label{fig:S2}
\end{figure}

Similar behavior is observed in the dynamics of the wave function $\ket{\psi}$ and coupling amplitudes $J_m$. We observe that the localized soliton-like wave packet travels along the array preserving its shape. The spatial distribution at the initial $t=0$ [Fig.~\ref{fig:S2}(a,d)], intermediate $t=\tau/2$ [Fig.~\ref{fig:S2}(b,e)], and final $t = \tau$ [Fig.~\ref{fig:S2}(c,f)] moments of time again can be fitted by the time evolution for a single site and single coupling link, respectively.

\subsection{B. Analytical fit of the solution}

To construct an initial guess for the soliton-like solution in our system, we utilize the following analytical formula based on our numerical simulations for the finite-size arrays. The real part $x_m$ of the Lax eigenvector is approximated by
\begin{eqnarray}
    x_m(\zeta) = A_N\exp{-2\Theta_p(\zeta^2,1/e)}\;,
\end{eqnarray}
where 
$\zeta = m-N(t-0.5)/\tau$
and $\Theta_p(p,q) = \sum_{m=0}^\infty p^m q^{m^2}$ is a partial theta function and 
$A_N = (19.1147)^{-1/2}$ is a normalization coefficient chosen such that $\sum_{m}x_m^2 = 1/2$.

The imaginary part of the Lax eigenvector $y_m$ is fitted by the approximate formula
\begin{eqnarray}
    y_m(\zeta) = \tanh(0.8 \zeta) (0.14853 + \exp[ -1.38 - 0.3 \zeta^2 - 0.08 \zeta^4 + 0.009 \zeta^6 - 
   0.00028 \zeta^8])\,.
\end{eqnarray}

Using this as an initial guess, the values $x_m(0)$ and $y_m(0)$ are further varied by the shooting method providing the transfer and satisfying normalization $\sum_{m}x_m^2 = 1/2$ and $\sum_{m}y_m^2 = 1/2$.

We also fitted the numerically found values $L_0$ for different lengths $N$ of the array by the expression
\begin{eqnarray}
    L_0(N)\tau = 0.58 + 0.42 N^{0.58}\;.
\end{eqnarray}

\begin{figure}[b]
    \centering
    \includegraphics[width=0.8\linewidth]{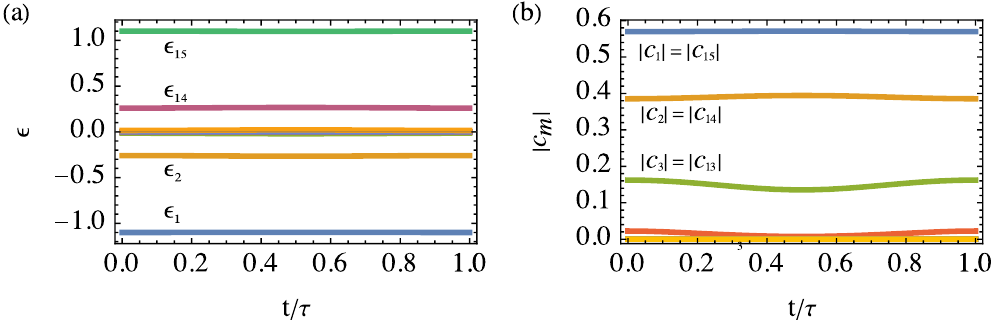}
    \caption{(a) Instantaneous spectrum of the Hamiltonian $\hat{H}(t)$. (b) Absolute values of the expansion coefficients $c_m$.}
    \label{fig:S3}
\end{figure}

\section{V. Non-adiabatic nature of the protocol and Aharonov–Anandan phase}

It should be stressed that the protocol of the state transfer which we study here is essentially non-adiabatic. To exemplify that, we diagonalize the Hamiltonian $\hat{H}(t)$ at each moment of time and plot its instantaneous spectrum in Fig.~\ref{fig:S3}(a). We observe that the eigenvalues remain practically constant in time. Due to the chiral symmetry of the problem, the spectrum is symmetric with respect to the zero energy (which is set by the eigenfrequency of a single qubit). Most of the eigenvalues are zero. This is due to the fact that most of the couplings in the array at each moment of time are zero. 

Next, having a solution for the wave function $\ket{\psi(t)}$ we project it onto the instantaneous eigenstates. We observe that the state of the system at each moment of time is a {\it superposition} of instantaneous eigenstates, and there is no single eigenstate which dominates. This brings us to the conclusion that the present protocol is not connected to the adiabatic limit and provides a distinct sort of physics.

In this light, it is surprising that there still exist  quantized invariants capturing this non-adiabatic evolution. We note that after the time $N\,\tau$ the soliton-like wave packet returns to its original position. During the propagation in the ring, the phase of its wave function jumps by $\phi=\pi$ due to the anti-periodic boundary condition. Therefore, we can introduce a vector $\ket{\tilde{\psi}}=e^{-i \phi t/(N\tau)}\ket{\psi}$ with a smooth dependence on $t$. Interpreting $t$ as a parameter of the system, we then evaluate the canonical Berry phase (which is also called in this context Aharonov-Anandan phase~\cite{Aharonov1987})
\begin{eqnarray}
\label{eqS:AAphase}
    \gamma = i\int_0^{N\tau}\bra{\tilde{\psi}}\partial_t\ket{\tilde{\psi}}dt = \phi + \int_0^{N\tau}\bra{\psi}\hat{H}\ket{\psi}dt\:.
\end{eqnarray}
In our case, the phase equals $\pi$ with high precision. This simple calculation suggests that there is a prospect to associate some nontrivial geometric phases with time-optimal non-adiabatic protocols.

\section{VI. Assessing bosonic and fermionic cases}

In the analysis above we presented the Hamiltonian in terms of the bosonic operators $\cra{m}$ using their commutation relations in the calculations. This representation builds on the fact that the superconducting transmon qubit can be described as quantum anharmonic oscillator obeying the Bose-Hubbard model
\begin{equation}    \hat{H}=\omega_0\hat{a}^\dagger\hat{a}+\frac{U}{2}\,(\hat{a}^\dagger)^2\hat{a}^2\:.
\end{equation}
The limit of $U\rightarrow 0$ corresponds to a quantum harmonic oscillator, while the limit of hardcore bosons $U\rightarrow\infty$ is equivalent to the idealized two-level system.

The cases with bosonic and fermionic statistics are clearly different. However, this distinction is revealed only in the many-body case. As long as we deal with the single-particle excitations, bosonic or fermionic statistics does not show up, and therefore our conclusions are equally valid both for the array of harmonic oscillators and two-level quantum systems. This justifies the terminology used in the manuscript which discusses the array of qubits, but exploits bosonic operators. To avoid confusion, most of the equations are written in the first quantization using the notations of bra and ket vectors and without using the specific statistics.

\section{VII. Numerical solution for the lattice of 10~000 qubits}

In this section, we demonstrate the potential of our method to treat truly large-scale quantum systems and provide the numerical solution for the lattice consisting of 10~000 qubits. We examine  two representative situations: (i) a finite lattice of 10 000 qubits with open boundary conditions, end-to-end transfer; (ii) a lattice of 10 000 qubits with anti-periodic boundary conditions, when the wave packet propagates keeping its shape.

\subsection{A. Open boundary conditions}
\begin{figure}[h!]
    \centering
    \includegraphics[width=0.8\linewidth]{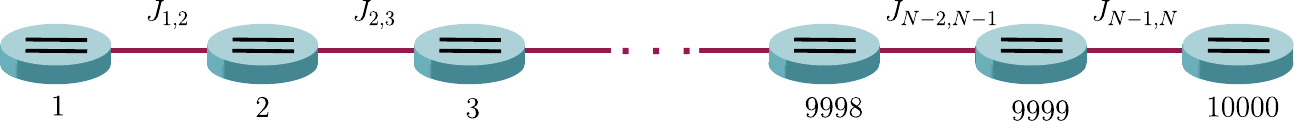}
    \caption{Sketch of the qubit array with open boundary conditions. We study the time-optimal transport of the excitation from qubit 1 to qubit 10 000.}
    \label{fig:OB10000}
\end{figure}

We consider a lattice of qubits with open boundary conditions shown in Fig.~\ref{fig:OB10000} and described by the tight-binding Hamiltonian $\hat{H}=\sqrt{2}\sum_{m=1}^{N-1} J_m(t)\,\hat{A}_{m,m+1}$. 
The real-valued couplings $J_m(t)$ are assumed to be switchable in time. The goal is to find a control protocol that transfers the state from $\ket{\psi_0}=\ket{1}$ to $\ket{\psi_1}=\ket{N}$ for $N=10~000$ within the minimal time.

\begin{figure}
    \centering
    \includegraphics[width=0.49\linewidth]{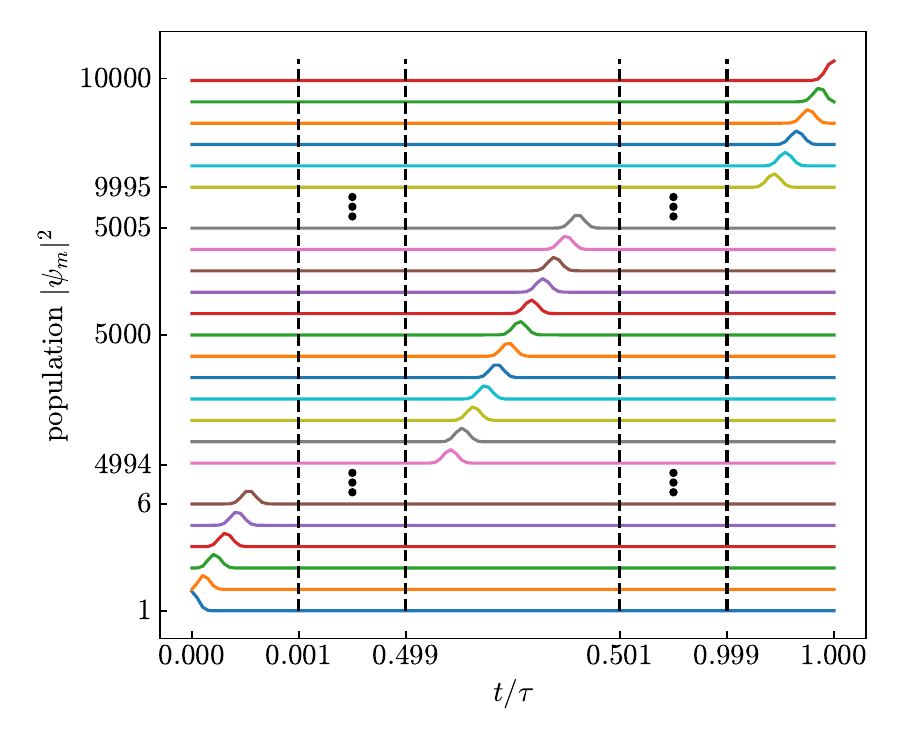}
    \includegraphics[width=0.49\linewidth]{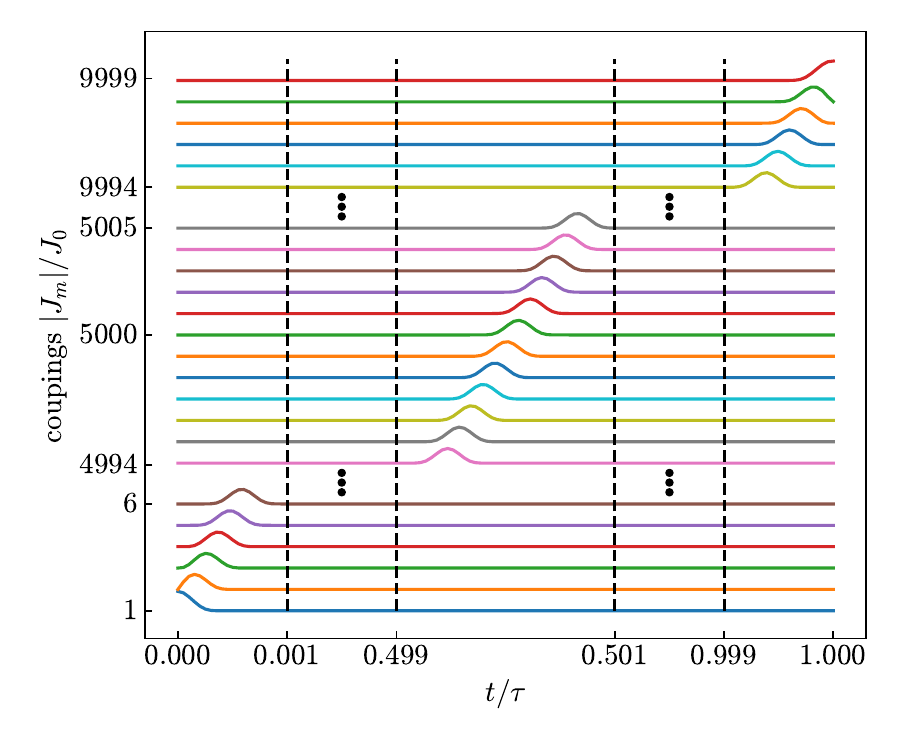}
    \caption{The state population $|\psi_m|^2$ is transferred as a localized wave packet (Left) under the time-optimal coupling amplitude control  $J_m (t)$ (Right).}
    \label{fig:evolution10000}
\end{figure}

Using the developed method, we have found the control protocol numerically. In Fig.~\ref{fig:evolution10000} we provide the evolution of the qubit population $|\psi_m|^2$ and the respective coupling control $J_m(t)$. The state travels as a localized wavepacket, and the transfer is controlled by the synchronously moving distribution of the couplings.

For the calculated solution, the state is transferred in a lattice of $N = 10~000$ qubits with the fidelity $F = 0.9994$, which is sufficiently close to unity (the fidelity can be improved further, but the protocol does not change much). The transfer time reads   $J_0 \tau = 11304.37262$. Combining this result with our calculations for smaller systems, we can approximate the transfer time in a lattice of $N$ qubits by the formula
\begin{equation}
J_0 \tau = 1.13031\,N + 1.27262\:.    
\end{equation}
On the other hand, in the absence of boundary effects, when the lattice is closed into the ring, the transfer takes a bit smaller time given by the expression
\begin{equation}
J_0 \tau'=1.13031\,(N-1)\:.    
\end{equation}
Reshaping of the wave packet close to the boundaries of the open-ended lattice takes an extra time given by the difference of the above expressions. The reshaping time reads 
$J_{0}\tau_b = 2.40293$.
These results provide an instance of the optimal control calculated for the lattice of 10 000 qubits.

For better reproducibility of the solution, we  provide the results of optimized values of $L_0 = 635025.6$
and the imaginary part of the Lax eigenvector $y_m(0)$ in Table~\ref{tab:y_OB_10000b}-\ref{tab:y_OB_10000e}.
\begin{table}[h!]
    \centering
    \begin{tabular}{|c|c|c|c|c|c|c|c|c|c|}
    \hline
        $m$ & 1 & 2 & 3 & 4 & 5 & 6 & 7 & 8 & 9 \\
        \hline
        $L_0 y_m/J_0$ & 0 & 0.707 & 0.4409 & 0.4066 & 0.39925 & 0.3976 & 0.3973 & 0.3973 & 0.3972\\
        \hline
    \end{tabular}
    \caption{First 9 values of the normalized imaginary part of the Lax eigenvector $y_m(0)$.}
    \label{tab:y_OB_10000b}
\end{table}

\begin{table}[h!]
    \centering
    \begin{tabular}{|c|c|c|c|c|c|c|c|c|}
    \hline
        $m$ & 10$\leq m\leq$ 9993 & 9994 & 9995 & 9996 & 9997 & 9998 & 9999 & 10000\\
        \hline
        $L_0 y_m/J_0$ & 0.3972 & 0.3972 & 0.3972 & 0.3971 & 0.3969 & 0.3959 & 0.3906 & 0.36175\\
        \hline
    \end{tabular}
    \caption{Intermediate and last 7 values of the normalized imaginary part of the Lax eigenvector $y_m(0)$.}
    \label{tab:y_OB_10000e}
\end{table}

\subsection{B. Periodic boundary conditions}
For the periodic boundary conditions and $N=10~000$, we use the solution from section~IV ($N=15$) as an initial guess. Since we have the functional approximation of the real and imaginary parts of the Lax eigenvector, the solution traveling as a localized wave packet and recovering itself after a shift by one lattice site is easily scalable to larger arrays. Specifically, here we provide a solution for $N=10~000$ with infidelity of one-site shift $1-F(\tau)\sim 10^{-10}$. The vector for the initial state is given by Table~\ref{tab:y_per} and the Lax eigenvalue is $L_0 = 63.496384$.

\begin{table}[h!]
    \centering
    \begin{tabular}{|c|c|c|c|c|c|}
    \hline
       $m $ & 0 & 1 & 2 & 3 & $m>3$ \\
       \hline
       $x_m $ & 0.59193 & 0.27299 & 0.01679 & 0.00003 & 0\\
       \hline
       $L_0 y_m$ & 0 & 0.63862 & 0.51327 & 0.45046 & 0.44895 \\
       \hline
       $L_0 y_{-m}$  & 0 & -0.63862 & -0.51327 & -0.45046 & -0.44895 \\
       \hline
    \end{tabular}
    \caption{First 4 values of the Lax eigenvector, where only $m\geq 0$ are shown since $x_m(0) = x_{-m}(0)$}
    \label{tab:y_per}
\end{table}

\begin{figure}[h!]
    \centering
    \includegraphics[width=0.6\linewidth]{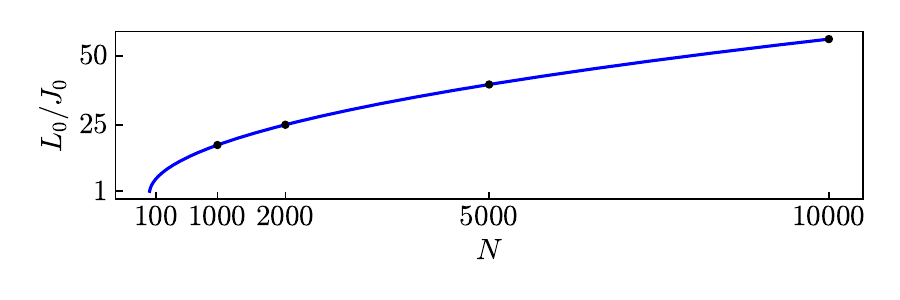}
    \caption{Scaling of the ratio $L_0/J_0$ with the size of the chain $N$. Black dots are exact numerical results, and a blue line is a fit.}
    \label{fig:L_0N}
\end{figure}

Finally, we highlight that both for periodic and open boundary conditions, the propagating wave packet has very similar shape. Therefore, it is expected that the two problems share the same invariant property. By computing the protocol for different $N$, we recover the scaling of the Lax eigenvalue $L_0/J_0 = \sqrt{0.3945+0.315526 N}$ (Fig.~\ref{fig:L_0N}) for both anti-periodic and open-ended chains.

\section{VIII. Time-optimal transfer in a lattice with constant couplings}

In this section, we consider the application of our method to a different physical system, which is a nearest-neighbor-coupled lattice of identical qubits with time-independent couplings $J_m$. We seek for such spatial distribution of these static couplings that minimizes the transfer time of the single-particle excitation between the states $\ket{1}$ and $\ket{N}$, given that the sum of squares of the couplings is fixed: $\sum_m\,J_m^2=J_0^2$.

An additional constraint making the couplings $J_m$ time-independent results in the cost functional of the form
%
\begin{equation}\label{eqs:Cost_С}    
\begin{split}
    S=\frac{1}{2}\int\limits_0^\tau\,\|\hat{H}\|^2\,dt+\int\limits_0^\tau\,\text{Tr}(\hat{H}\hat{D})\,dt+
\int\limits_0^\tau\, \mathrm{Tr}\left(\hat{R}\,(\partial_t\hat{\rho}+i[\hat{H},\hat{\rho}])\right)\;dt\\
+\sqrt{2}\int\limits_0^\tau\,\sum_m \mu_m \left(\mathrm{Tr}\left(\hat{H}\hat{A}_{m,m+1}\right)-\sqrt{2}J_m\right)\,dt\,.
\end{split}
\end{equation}
This functional explicitly includes the constraints
\begin{gather}
\mathrm{Tr}\left(\hat{H}\hat{D}_k\right) = 0\;,\label{eq:emc4}\\
\partial_t\hat{\rho} +i[\hat{H},\hat{\rho}]=0\;,\label{eq:emc2}\\
\mathrm{Tr}\left(\hat{H}\hat{A}_{m,m+1}\right)=\sqrt{2}J_m\label{eq:emc5}
\end{gather}
obtained by varying with respect to the Lagrange multipliers $\lambda_k$, $\hat{R}$ and $\mu_m$. Variations with respect to the dynamical variables $\hat{H}$ and $\hat{\rho}$ yield
\begin{eqnarray}
\label{eq:emc1}
 &\hat{H}+\hat{D}+\sqrt{2}\sum_m\mu_m\hat{A}_{m,m+1} +i[\hat{\rho},\hat{R}] = 0\;,\\
\label{eq:emc3}
&\partial_t\hat{R} +i[\hat{H},\hat{R}]=0\;.
\end{eqnarray}
In these equations, the couplings $J_m$ are constant in time, but their specific values are not fixed. Therefore, the choice of $J_m$ should also extremize the cost functional via $\partial S/\partial J_m=0$. This requirement yields:
\begin{equation}\label{eq:emc6}
\int_0^\tau \mu_m dt = 0\,.
\end{equation}
We define the Lax operator in the same way as previously:
\begin{equation}
\hat{L} = i[\hat{R},\hat{\rho}]\:.    
\end{equation}
Equation~\eqref{eq:emc1} then yields
\begin{equation}\label{eq:LaxNew2}
    \hat{L} = \hat{H} + \hat{D} + \sqrt{2}\sum_m \mu_m \hat{A}_{m,m+1}\;.
\end{equation}
We now define the eigenvectors of the Lax operator
\begin{eqnarray}
    \hat{L}\ket{\alpha} = L_0\ket{\alpha}\;,\\
    \hat{L}\ket{\beta} = -L_0\ket{\beta}\;.
\end{eqnarray}
Similarly to the section II, the Hamiltonian $\hat{H}$ satisfies chiral symmetry $\hat{H}\hat{\sigma}+\hat{\sigma}\hat{H} = 0$ and the evolution of the chiral part of the Lax operator $\hat{L}_A$ is decoupled, resulting in the symmetry of the spectrum $\ket{\beta} = \hat{\sigma}\ket{\alpha}$. We use $\hat{Q} = \sum_m i^{m-1}\ket{m}\bra{m}$ unitary transformation to rewrite the chiral part of the Lax operator as
\begin{eqnarray}
    \hat{L}^Q_A = L_0(\ket{a}\bra{a}-\ket{a^*}\bra{a^*})\:.
\end{eqnarray}
The Lax eigenvectors evolve according to the Schr\"odinger equation $i\partial_t\ket{a}=\hat{H}^Q\ket{a}$, which in this basis reads
\begin{eqnarray}
    \partial_t a_m = - J_m a_{m+1} + J_{m-1}a_{m-1}\;.
\end{eqnarray}
The real and imaginary parts $\ket{a} = \ket{x}+i\ket{y}$ are decoupled and follow the same equation
\begin{eqnarray}
    \partial_t x_m = - J_m x_{m+1} + J_{m-1}x_{m-1}\;,\label{eq:Evolx}\\
    \partial_t y_m = - J_m y_{m+1} + J_{m-1}y_{m-1}\;.\label{eq:Evoly}
\end{eqnarray}
In addition to that, the evolution must also satisfy Eq.~\eqref{eq:emc6}. By projecting Eq.~\eqref{eq:LaxNew2} onto $\hat{A}_{m,m+1}^Q$, we obtain
\begin{eqnarray*}
    \sqrt{2}\,(\mu_m + J_m) &=& \mathrm{Tr}\left(\hat{L}^Q_A\hat{A}_{m,m+1}^Q\right)\;,\\
    \mu_m + J_m &=& iL_0\,(a_m a_{m+1}^* - a_m^* a_{m+1})\;,\\
    \mu_m + J_m &=& 2 L_0(x_m y_{m+1}-x_{m+1}y_{m})\;.
\end{eqnarray*}
Hence, the integral equation~\eqref{eq:emc6} takes the form
\begin{equation}\label{eq:IntegralEq}
2 L_0\,\int_0^\tau (x_m y_{m+1}-x_{m+1}y_{m})\,dt  = J_m \tau\;.
\end{equation}
Equations~\eqref{eq:Evolx},\eqref{eq:Evoly},\eqref{eq:IntegralEq} are the governing equations in our problem.

Since the Hamiltonian is constant in time, the vectors $\ket{x}$ and $\ket{y}$ can be expanded in the basis of the Hamiltonian eigenstates $\ket{k^m}$, and their temporal evolution is readily computed. If one chooses some specific pattern of the couplings $J_m$, this defines the eigenmodes of the Hamiltonian. The initial condition for the wave function $x_m(0)=\delta_{m,1}/\sqrt{2}$ determines the expansion coefficients for the $\ket{x}$ vector. Once the evolution of $\ket{x}$ is found, the expansion coefficients $c_p$ for the vector $\ket{y}$ are defined from Eq.~\eqref{eq:IntegralEq}, which is converted to the linear system of equations of the form
$\sum_p F_{m,p} c_p = J_m$. Of course, the chosen set of $J_m$ should provide the transfer, satisfying the boundary condition $x_m(\tau) = \delta_{m,N}/\sqrt{2}$.




To derive the solution of Eqs.~\eqref{eq:Evolx},\eqref{eq:Evoly},\eqref{eq:IntegralEq}, we exploit the symmetries and general properties of the system. For clarity, the rest of this section is written in the original representation with the real couplings.

First, we note that all couplings in the lattice must be nonzero $J_m\neq 0$, otherwise there is no way for the excitation to be transferred from one part of the lattice to another. In the other words, the lattice of $N$ sites is connected. For a connected system with nearest-neighbor couplings, the eigenvector $\ket{k^m}$ corresponding to the given eigenvalue $E_m$ is uniquely determined. Indeed, the eigenvalue equation reads
\begin{eqnarray}
    \hat{H}\ket{k^m} = E_m\ket{k^m}\;,\\
    J_{p}k_{p+1}^m + J_{p-1}k_{p-1}^m = E_m k_p^m\;,
\end{eqnarray}
allowing unique definition of all coefficients $k_p^m$ via $E_m$ and $k_1^m$. The latter is fixed by the normalization, so the spectrum of the lattice is non-degenerate.

Second, the initial and target states are related via the reflection symmetry $\ket{N}=\hat{\Gamma}\ket{1}$, where
\begin{equation}\label{eq:ReflOperator}
\hat{\Gamma} = \sum\limits_{m=1}^{N} \ket{m}\bra{N-m+1}
\end{equation}
is the reflection operator, satisfying the property $\hat{\Gamma}^2=\hat{I}$. By construction, we seek such evolution that yields
\begin{equation}\label{eq:transfer}
\hat{U}(\tau)\ket{1} = e^{i\phi}\ket{N}\:,    
\end{equation}
where
\begin{equation}\label{eq:EvolOperator}
\hat{U}(\tau)=e^{-i\hat{H}\tau}    
\end{equation}
and $\phi$ is the evolution phase. In the representation with the real-valued couplings the Hamiltonian matrix is symmetric: $\hat{H}^T=\hat{H}$. Equation~\eqref{eq:EvolOperator} then guarantees that $[\hat{U},\hat{H}]=0$ and $\hat{U}=\hat{U}^T$. Since in addition $\hat{U}$ is unitary,
\begin{equation}\label{eq:Uprop}
\hat{U}^{-1}=\hat{U}^\dagger=\hat{U}^*\:.
\end{equation}
Multiplying Eq.~\eqref{eq:transfer} by $\hat{U}^{-1}$ and applying the above property, we recover
\begin{equation}\label{eq:transfer}
\hat{U}(\tau)\ket{N} = e^{i\phi}\ket{1}\:.    
\end{equation}
Thus, the same protocol which transfers the state from $\ket{1}$ to $\ket{N}$ also transfers the state backward from $\ket{N}$ to $\ket{1}$, while $2\phi$ is the geometric phase acquired during such an evolution.

Moreover, we can also compute the entire matrix $\hat{U}(\tau)$. We use that the Hamiltonian $\hat{H}$ is the tri-diagonal matrix commuting with the evolution operator:
\begin{eqnarray}
    \hat{H}\ket{m} &=& J_{m}\ket{m+1} + J_{m-1}\ket{m-1}\;,\\
    \hat{U}\hat{H}\ket{m} &=& J_{m}\hat{U}\ket{m+1} + J_{m-1}\hat{U}\ket{m-1}\;,\\
    \hat{H}\hat{U}\ket{m} &=&  J_{m}\hat{U}\ket{m+1}+ J_{m-1}\hat{U}\ket{m-1}\;.\label{eq:recursive}
\end{eqnarray}
This is the recursive expression, where we know $\hat{U}\ket{1}$, $\hat{U}\ket{N}$, while $\ket{0}\equiv0$ and $\ket{N+1}\equiv0$. Applying Eq.~\eqref{eq:recursive} for $m=1$ and $m=N$, we find:
\begin{gather}
    J_1\hat{U}\ket{2} = e^{i\phi}J_{N-1}\ket{N-1}\:,\\
    U_{N-1,2} = e^{i\phi}J_{N-1}/J_1\;.\\
    J_{N-1}\,\hat{U}\ket{N-1} = J_1\,e^{i\phi}\ket{2}\:,\\
    U_{2,N-1}=e^{i\phi} J_1/J_{N-1}\:.
\end{gather}
However, $\hat{U}$ is symmetric and therefore $U_{2,N-1}=U_{N-1,2}$, which yields $|J_1| = |J_{N-1}|$. Since we are interested in the positive coupling constants, this means
\begin{gather}
    J_1=J_{N-1}\:,\\
    \hat{U}\ket{2}=e^{i\phi}\ket{N-1}\:,\\
    \hat{U}\ket{N-1}=e^{i\phi}\ket{2}\:.
\end{gather}
%
%
On the next step, we compute
\begin{gather}
J_2\hat{U}\ket{3}=\hat{H}\hat{U}\ket{2}-J_1\hat{U}\ket{1}=\hat{H}e^{i\phi}\ket{N-1}-J_1e^{i\phi}\ket{N}\notag\\
=e^{i\phi}\,\left(J_{N-1}\ket{N}+J_{N-2}\ket{N-2}-J_1\ket{N}\right)=e^{i\phi}\,J_{N-2}\,\ket{N-2}\:.
\end{gather}
In the same spirit, we recover
\begin{gather}
J_{N-2}\hat{U}\ket{N-2}=\hat{H}\hat{U}\ket{N-1}-J_{N-1}\hat{U}\ket{N}=\hat{H}e^{i\phi}\ket{2}-J_{N-1}e^{i\phi}\ket{1}\\
=e^{i\phi}\,\left(J_1\ket{1}+J_3\ket{3}-J_{N-1}\ket{1}\right)=e^{i\phi}J_3\ket{3}\:.
\end{gather}
This leads to the similar set of relations:
\begin{gather}
J_2=J_{N-2}\:,\\
\hat{U}\ket{3}=e^{i\phi}\ket{N-2}\:,\\
\hat{U}\ket{N-2}=e^{i\phi}\ket{3}
\end{gather}
with an iterative generalization
\begin{gather}
J_m=J_{N-m}\:,\label{eq:CouplingPat}\\
\hat{U}\ket{m+1}=e^{i\phi}\ket{N-m}\:,\\
\hat{U}\ket{N-m}=e^{i\phi}\ket{m+1}
\end{gather}
for $m=1,2,\dots(N-1)$. This analysis brings us to the conclusion that the pattern of the couplings in the tight-binding lattice under study must necessarily be symmetric, while the evolution operator
\begin{equation}\label{eq:EvolTau}
\hat{U}(\tau)=e^{i\phi}\,\hat{\Gamma}\:.
\end{equation}
The direct consequence of Eq.~\eqref{eq:EvolTau} is
\begin{equation}
\hat{U}(2\tau)=\hat{U}^2(\tau)=e^{2i\phi}\,\hat{I}\:.
\end{equation}
This means that the evolution of the {\it arbitrary} initial quantum state is periodic with the period $T=2\tau$. The above derivation complements and generalizes that in the End Matter section.


In turn, the periodicity of the evolution leads to the quantization of the eigenenergies
$E_m =(-\theta+2n_m\pi)/2\tau$, where $n_m$ is an integer number different for each $m$, and $\theta=2\phi$ is a constant equal either to $0$ or $\pi$ to ensure chiral-symmetric spectrum with $E_{m}=-E_{N-m+1}$. We enumerate the energy levels such that for $n>m$ the energies are related as $E_n>E_m$. We introduce the differences  $g_m = n_{m+1}-n_m\geq1$ and express
\begin{eqnarray}
    E_m =\frac{-\theta+2n_m\pi}{2\tau}=\pi\,\dfrac{n_m-n_{N-m+1}}{2\tau} = \dfrac{\pi}{2\tau}\text{sign}(m-(N+1)/2) \sum_{k=\min(m,N-m+1)}^{\max(m,N-m+1)-1} g_k\;.
\end{eqnarray}

We now compute the coupling resource
\begin{eqnarray}
    J_0\tau =\tau\,\sqrt{\frac{\mathrm{Tr}\hat{H}^2}{2}} = \dfrac{\tau}{\sqrt{2}}\sqrt{\sum_m E_m^2} 
    = \dfrac{\pi}{2\sqrt{2}}\sqrt{\sum_{m=1}^N \left(\sum_{k} g_k\right)^2}\;,
\end{eqnarray}
where we can estimate
\begin{eqnarray}
    \left|\sum_{k=\min(m,N-m+1)}^{\max(m,N-m+1)-1} g_k\right| \geq |2m-N-1|\;,
\end{eqnarray}
with equality for $g_k = 1$. The sum $\sum_{m=1}^N (2m-N-1)^2=N(N^2-1)/3$, and the bound on the coupling resource reads
\begin{eqnarray}
    J_0\tau \geq \pi\sqrt{N(N^2-1)/24}\;.
\end{eqnarray}
with the lower bound reached when all $g_k=1$. Inspecting the literature~\cite{Christandl2004}, we recover that the equidistant energy spectrum ($g_k = 1$) and the above bond on the coupling resource correspond to the coupling pattern $J_m=B/2\sqrt{m(N-m)}$, which is known as perfect transfer strategy. This strategy can be applied to the lattices of arbitrary length $N$, while our analysis above proves the global optimality of such transfer protocol.

\section{IX. Comparison with the other optimal control methods}

In this section, we discuss how the technique developed in this paper compares to the other methods for computing quantum optimal control, including  GRAPE, CRAB, Krotov, and GOAT.

The first important difference is that all these methods are fixed-time fidelity maximizers: given the duration of the protocol $\tau$ and the control ansatz, they search for the control that maximizes the fidelity. Therefore, to find a time-optimal protocol with unit fidelity, double-target optimization is required. For example, this may require wrapping the fidelity maximization in an outer loop over $\tau$ (or over an overall resource $J_0$), re-running the entire optimization at each trial value, and searching for the scenario when the fidelity reaches unity. Such nested search is computationally expensive, offers no guarantee of convergence to the minimal time, and often tends to produce non-smooth, iteration-count-dependent solutions even when it succeeds. 

The quantum brachistochrone method, by contrast, treats $\tau$ itself as a variable extremized directly in the variational problem, with unit fidelity imposed as a hard boundary condition rather than an outer-loop target. Specifically, time-optimality is imposed by the analytical equations of motion derived from the variational principle, while the fidelity is maximized numerically by varying the initial values in such a way that the boundary conditions are satisfied. This provides the key difference in the structure of the optimization problem.  

To highlight this difference, we consider a conceptually simple example of a 3-qubit lattice with a Hamiltonian
\begin{eqnarray}
    \hat{H}_{3q} = J_1(t) (\ket{1}\bra{2} + \ket{2}\bra{1}) + J_2(t) (\ket{2}\bra{3} + \ket{3}\bra{2})\;,
\end{eqnarray}
where the time-dependent couplings are defined as $J_1(t) = J_0\cos(\Omega(t))$ and $J_2(t) = J_0\sin(\Omega(t))$ with unknown function $\Omega(t)$ to satisfy the constraint 
\begin{eqnarray}
    J_1^2+J_2^2 = J_0^2\;.
\end{eqnarray}

The quantum brachistochrone technique allows one to find the time-optimal solution to this problem analytically~\cite{Chernova2025}:
\begin{eqnarray}
    \Omega(t) = J_0 t/\sqrt{3}\;,\\
    J_0\tau =\sqrt{3} \pi/2\;,
\end{eqnarray}
which ensures the transport of the initial state $\ket{\psi(0)}=\ket{1}$ to $\ket{\psi(\tau)}=\ket{3}$.

\begin{figure}[h!]
    \centering
    \includegraphics[width=0.5\linewidth]{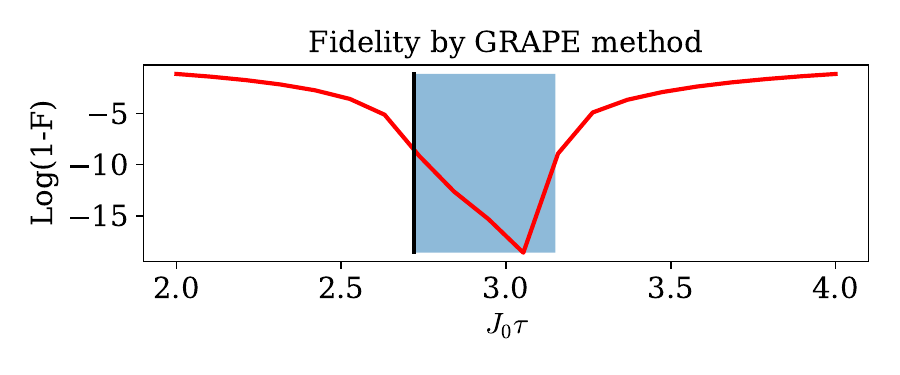}
    \caption{The values of transfer infidelity $1-F$ on a logarithmic scale for different $J_0\tau$ values by the GRAPE method. The vertical black line corresponds to a time-optimal solution $J_0\tau \simeq 2.7207$. Shaded area indicates the set of solutions enabling practically perfect transfer.}
    \label{fig:grape}
\end{figure}

Further, we use the GRAPE method to find the control $\Omega(t)$, using an analytical solution as an initial guess $\Omega_i(t) =\pi t/2$, and further minimize the parameter $J_0\tau$. To simplify numerical evaluation, we fix $\tau = 1$ and perform single-target optimization of the transport fidelity $F = |\bra{3}\ket{\psi(\tau)}|^2$ for a fixed coupling amplitude $J_0$. Values of infidelity $1-F$ evaluated in the range $J_0\in[2,4]$ are shown in Fig.~\ref{fig:grape}. We observe that the GRAPE method finds an appropriate transfer strategy for various values of $J_0$. Finding the minimal $J_0$ in that set is a nontrivial task. Furthermore, the protocol with fidelity $F=1$ and minimal time may not be found by the GRAPE algorithm, as it generally produces non-smooth solutions.

This brings us to the conclusion that the combination of quantum brachistochrone method and Lax pair approach developed in this manuscript is a tool specifically tailored to search for time-optimal dynamics and to accurately evaluate quantum speed limits in various problems. In such situation, our methodology streamlines the calculations compared to the general purpose quantum optimal control methods. In addition, our methodology provides an analytical insight in many instances.



\end{document}